\documentclass[11pt, epsfig]{article}                            
\usepackage{epsfig}       
\usepackage{amsfonts}    
\usepackage{amscd,verbatim}
\usepackage[all]{xy}
\usepackage{graphicx}
\usepackage{amsmath}    
\usepackage[psamsfonts]{amssymb}
\usepackage{amsthm}
\usepackage{euscript}

\newskip\humongous \humongous=0pt plus 1000pt minus 1000pt

\newif\ifdtup

\catcode`\@=11
 
\@addtoreset{equation}{section}
\def\theequation{\thesection.\arabic{equation}}
 
\def\@normalsize{\@setsize\normalsize{15pt}\xiipt\@xiipt
\abovedisplayskip 14pt plus3pt minus3pt%
\belowdisplayskip \abovedisplayskip
\abovedisplayshortskip \z@ plus3pt%
\belowdisplayshortskip 7pt plus3.5pt minus0pt}
 
\def\small{\@setsize\small{13.6pt}\xipt\@xipt
\abovedisplayskip 13pt plus3pt minus3pt%
\belowdisplayskip \abovedisplayskip
\abovedisplayshortskip \z@ plus3pt%
\belowdisplayshortskip 7pt plus3.5pt minus0pt
\def\@listi{\parsep 4.5pt plus 2pt minus 1pt
     \itemsep \parsep
     \topsep 9pt plus 3pt minus 3pt}}
 
\relax

\catcode`@=12
 
\catcode`\@=11
 
\def\section{\@startsection{section}{1}{\z@}{3.5ex plus 1ex minus
   .2ex}{2.3ex plus .2ex}{\large\bf}}
 
\def\thesection{\arabic{section}}    
\def\thesubsection{\arabic{section}.\arabic{subsection}}

\def\appendix{\setcounter{section}{0}
 \def\thesection{Appendix \Alph{section}}
 \def\thesubsection{\Alph{section}.\arabic{subsection}}
 \def\theequation{\Alph{section}.\arabic{equation}}}
\def\YGrule{0.4}   % line thickness in unit of pt
\def\YGbox{6.5}    % box size in unit of pt
\def\SymBoxes#1#2#3#4{\newdimen\un@t \un@t#3%
\raisebox{#1}{\rule{#2\un@t}{#4}\hskip-#2\un@t% lower horizontal
\@tempdimb\un@t \advance\@tempdimb by-#4\@tempcntb#2\relax%
\@whilenum{\@tempcntb>0}\do{%                         % #2 vertical lines
\rule{#4}{\un@t}\hskip\@tempdimb \advance\@tempcntb by\m@ne}%
\hskip-#2\un@t \rule[\un@t]{#2\un@t}{#4}%
\rule[\un@t]{#4}{#4}\hskip-#4%             % upper horizontal line
\rule{#4}{\un@t}}\hskip-#4}                % rightest vertical line
\def\Young{\@ifnextchar[{\@Young}{\@Young[0]}}
\def\@Young[#1]#2{\newdimen\YG@unit \YG@unit\YGbox pt%
\newdimen\h@ight \h@ight#1\YG@unit \@tempcnta-1\relax
\@tfor\c@ount:=#2\do{\advance\@tempcnta by\@ne}% count the number of rows
\@tempdima\@tempcnta\YG@unit%
\advance\h@ight by\@tempdima\relax     % compute the height of the top row
\@tfor\c@ount:=#2\do{\SymBoxes{\h@ight}{\c@ount}{\YG@unit}{\YGrule pt}%
\@tempdima-\c@ount\YG@unit \hskip\@tempdima%
\advance \h@ight by -\YG@unit}         % Draw the Tableaux
\@tempdima\YG@unit \multiply\@tempdima by\@car#2\@nil %
\hskip\@tempdima}                      % hskip by the length of the top row
\def\YoungTab{\@ifnextchar[{\@YoungIdx}{\@YoungIdx[0]}}
\def\@YoungIdx[#1]{\@ifnextchar[{\@iYoungIdx[#1]}{\@iYoungIdx[#1][\@empty]}}
\def\@iYoungIdx[#1][#2]#3{%
\newdimen\YG@unit \YG@unit\YGbox pt\newdimen\YG@rule \YG@rule \YGrule pt
\newcount\c@ount \c@ount\z@ \newdimen\skip@wd \unitlength\@ne pt
\newdimen\h@ight \h@ight#1\YG@unit \@tempcnta\m@ne\relax
\@tfor\d@um:=#3\do{\advance\@tempcnta by\@ne}% count the number of rows
\@tempdima\@tempcnta\YG@unit%
\advance\h@ight by\@tempdima\relax%  % compute the height of the top row
\@tfor\@idxlist:=#3\do{%             % routine to draw the indexed Tableaux
\@tempcnta\z@\hskip.5\YG@rule\relax 
\@for\@idx:=\@idxlist\do{%           % place the indices of the row first
\raisebox{\h@ight}{\makebox(\YGbox,\YGbox){#2$\@idx$}}
\advance\@tempcnta by\@ne}\hskip-.5\YG@rule% 
\@tempdima-\@tempcnta\YG@unit \hskip\@tempdima%
\ifnum\c@ount=\z@ \skip@wd-\@tempdima\fi \relax% record the top row width
\SymBoxes{\h@ight}{\@tempcnta}{\YG@unit}{\YG@rule}%
\hskip\@tempdima \advance\h@ight by -\YG@unit
\advance\c@ount by\@ne}%             % end of the routine
\hskip\skip@wd}                      % hskip by the length of the top row
\newcommand{\hsp}{,\hspace{.3cm}}   
\newcommand{\gnr}[1]{\ensuremath{\langle #1\rangle}}

\newcommand{\pic}{\hspace{-.1cm}}
\newcommand{\ptimes}{\hspace{-.08cm}\times\hspace{-.08cm}}
\renewcommand{\a}{\ensuremath{\alpha}}
\renewcommand{\b}{\ensuremath{\beta}}
\newcommand{\hcup}{\ensuremath{\hspace{-.14cm}\stackrel{H}{\rightarrow}\hspace{-.12cm}}}  
\newcommand{\tha}{\ensuremath{\theta_\a}}  
\newcommand{\thb}{\ensuremath{\theta_\b}}  
\newcommand{\hsa}{\ensuremath{\widehat {S}^1_a}}

\newcommand{\hsd}{\ensuremath{\widehat {S}^1_d}}   
\newcommand{\R}{\ensuremath{\mathbb R}}
\newcommand{\rpn}{\ensuremath{{\mathbb R {\text{P}}^n}}}
\newcommand{\rpt}{\ensuremath{{\mathbb R {\text{P}}^3}}}
\newcommand{\rps}{\ensuremath{{\mathbb R {\text{P}}^7}}}
\newcommand{\rp}{\ensuremath{{\mathbb R {\text{P}}}}}
\newcommand{\cp}{\ensuremath{{\mathbb C {\text{P}}}}}

\newcommand{\C}{\ensuremath{\mathbb C}}

\def\H{{\text H}}

\newcommand{\N}{\ensuremath{\mathcal N}}

\newcommand{\claim}{\noindent {\bf Claim.\ }}

\begin{document}
%\begin{letter}{~}

%%%%%%Define some new commands and  macros
\newcommand{\beq}{\begin{equation}}
\newcommand{\eeq}{\end{equation}}
\newcommand{\bea}{\begin{eqnarray}}
\newcommand{\eea}{\end{eqnarray}}
\newcommand{\beas}{\begin{eqnarray*}}
\newcommand{\eeas}{\end{eqnarray*}}
\newcommand{\defi}{\stackrel{\rm def}{=}}
\newcommand{\non}{\nonumber}
\newcommand{\bquo}{\begin{quote}}
\newcommand{\enqu}{\end{quote}}
%%%%%%%%%%%%%%%%%%%%%%%%%%%%%%%%%% definitions
\def\de{\partial}
\def\Tr{ \hbox{\rm Tr}}
\def\const{\hbox {\rm const.}}
\def\o{\over}
\def\im{\hbox{\rm Im}}
\def\re{\hbox{\rm Re}}
\def\bra{\langle}\def\ket{\rangle}
\def\Arg{\hbox {\rm Arg}}
\def\Re{\hbox {\rm Re}}
\def\Im{\hbox {\rm Im}}
\def\diag{\hbox{\rm \footnotesize{diag}}}
\def\even{\hbox{\rm \footnotesize{even}}}
\def\odd{\hbox{\rm \footnotesize{odd}}}
\def\longvert{{\rule[-2mm]{0.1mm}{7mm}}\,}
\def\a{\alpha}
\def\dag{{}^{\dagger}}
\def\tq{{\widetilde q}}
\def\p{{}^{\prime}}
\def\W{W}
\def\N{{\cal N}}
\newcommand{\Z}{\ensuremath{\mathbb Z}}
\begin{titlepage}
\begin{flushright}
ULB-TH/05-02\\
%ADP-04-???/????\hsp
%MPP-2004-??\hsp
hep-th/0501110\\
\end{flushright}
%\bigskip
%\bigskip
\def\thefootnote{\fnsymbol{footnote}}

% =========================================================================

\begin{center}
{\Large  {\bf  Flux compactifications on projective spaces\\[0.3cm] 
and the S-duality puzzle}}
\end{center}
\bigskip

\begin{center}
{\large  Peter Bouwknegt\footnote{E-Mail: peter.bouwknegt@anu.edu.au} $^{(1,2)}$ , 
Jarah Evslin\footnote{E-Mail: jevslin@ulb.ac.be} $^{(5,6)}$,\\
Branislav Jur\v co\footnote{E-Mail: jurco@theorie.physik.uni-muenchen.de} $^{(7,8)}$, Varghese 
Mathai\footnote{E-Mail: mathai.varghese@adelaide.edu.au} $^{(4)}$, Hisham Sati\footnote{E-Mail: 
hisham.sati@adelaide.edu.au} $^{(3,4)}$
 \vskip 0.10cm
 }
\end{center}

\renewcommand{\thefootnote}{\arabic{footnote}}
\bigskip

\begin{center}
{\it   \footnotesize
Department of Theoretical Physics, Research School of 
Physical Sciences and Engineering $^{(1)}$\\
Department of Mathematics,
Mathematical Sciences Institute $^{(2)}$\\
The Australian National University, Canberra,
ACT~0200, Australia
\vspace{.3cm}\\

\noindent
Department of Physics and Mathematical Physics $^{(3)}$\\
Department of Pure Mathematics $^{(4)}$\\
University of Adelaide, Adelaide, SA 5005, Australia
\vspace{.3cm}\\

\noindent
International Solvay Institutes $^{(5)}$,\\
Physique Th\'eorique et Math\'ematique, Universit\'e Libre de Bruxelles $^{(6)}$,\\
C.P. 231, B-1050, Bruxelles, Belgium 
\vspace{.3 cm}\\

\noindent
Max-Planck-Institut fur Physik, Fohringer Ring 6, D-80805 $^{(7)}$\\
Sektion Physik, Universitat Munchen, Theresienstr. 37, D-80333 $^{(8)}$\\
Munchen, Germany

}
\end {center}

\bigskip
\bigskip
\noindent  
{\bf Abstract:}

We derive a formula for D3-brane charge on a compact spacetime, which includes torsion corrections to the tadpole cancellation condition.  We use this to classify D-branes and RR fluxes in type II string theory on $\rpt\times\rp^{2k+1}\times S^{6-2k}$ with torsion $H$-flux and to demonstrate the conjectured T-duality to $S^3\times S^{2k+1}\times S^{6-2k}$ with no flux.  When $k=1$, $H\neq 0$ and so the K-theory that classifies fluxes is twisted.  When $k=2$ the square of the $H$-flux yields an S-dual Freed-Witten anomaly which is canceled by a D3-brane insertion that ruins the dual K-theory flux classification.  When $k=3$ the cube of $H$ is nontrivial and so the D3 insertion may itself be inconsistent and the compactification unphysical.  Along the way we provide a physical interpretation for the AHSS in terms of the boundaries of branes within branes.%To perform the T-duality we need to add a globally-defined B-field which combines with a RR connection in a new Freed-Witten anomaly that cancels the charge of the insertions.

\vfill  
 
\end{titlepage}

\bigskip

\hfill{}
\bigskip
\setcounter{footnote}{0}

% =========================================================================
\section{Introduction }

\subsection{What is classified by twisted K-theory?}

If we compactify type II string theory on a compact manifold the
consistency of the D-brane partition functions implies that the
Ramond-Ramond field strengths, which we write locally as
$G_p=dC_{p-1}$, are quantized.  This may lead us to believe that if
we are not interested in changes by globally
defined connections $C_{p-1}$ then the $G_p$ are classified by
integral cohomology.

However several authors \cite{MM,Witb,BMKtheory} have suggested that instead RR
field strengths are classified by twisted K-theory, which is a quotient of a subset of integral cohomology.  For example, all of the field strengths that are in this subset satisfy
\beq
d_3 G_p=(Sq^3+H\cup) G_p=0 \,, \label{speciale}
\eeq
where $Sq^3$ is an operator that takes torsion $p$-classes to torsion $(p+3)$-classes\footnote{We do not 
assume that the reader is familiar with the Steenrod squares $Sq^i$.  However a crash course may be found in Ref.~\cite{DMW}.}.  In addition twisted K-theory identifies cohomology classes that differ by an element in the image of $d_3$, which is just another consequence of forgetting the globally defined connections \cite{EvsFlux}.

To see what is so special about the subset (\ref{speciale}), we turn our attention to the classical limit, 
type II supergravity. This means that we forget the quantization condition, so we are now looking at real 
cohomology and the torsion is gone.  In particular we can no longer see the $Sq^3$ term. This theory has 
RR potentials $C_{p-1}$, a Neveu-Schwarz (NS) 3-form $H$ and a peculiar gauge-invariance
\beq
C_{p-1}\longrightarrow C_{p-1}+d\Lambda_{p-2}+H\wedge \Lambda_{p-4} \,, \label{trasformazioni}
\eeq
for any set of forms $\Lambda_k$.  This means that there are two natural field strengths \cite{Mar}
\beq
G_{p}=dC_{p-1}\,\hspace{.5cm}\textup{and}\hspace{.5cm}F_p=dC_{p-1}+H\wedge C_{p-3}\,,
\eeq
of which $G_p$ is closed and $F_p$ is gauge-invariant.  As in QED, we introduce charges as violations of the Bianchi identity
\beq
Q_{\textup\footnotesize{{\textup{D}}(8-p)}}=ddC_{p-1}=dF_p-d(H\wedge C_{p-3})=dF_p+H\wedge F_{p-2}\,,
\eeq
where in the last step we have used the fact that $H\wedge H$ vanishes classically, although in the quantum theory it may have a torsion contribution.  Note that this notion of D-brane charge is not the notion of D-brane charge of, for example, Ref.~\cite{GKP}.  There the authors define D-brane charge to instead be $dF$, and refer to $dG$ as the charge contribution from local sources.

The classical limit of the condition (\ref{speciale}) is $H\wedge G=0$ but instead classical supergravity yields
\beq
H\wedge G_{p-2}=H\wedge F_{p-2}=Q_{\textup\footnotesize{{\textup{D}}(8-p)}}-dF_p\,. \label{meglio}
\eeq
Thus twisted K-theory seems to classify only fluxes in the subset of
configurations for which the right hand side of Eq.~(\ref{meglio})
vanishes.  There are many consistent string backgrounds that do not satisfy the condition (\ref{speciale}), these correspond to cohomology classes but not to K-theory classes. That is, K-theory classifies only those configurations in which all of the
branes are sources for the gauge-invariant field strength $F_p$ and not
branes created from $H\wedge F$, such as those constructed during
Hanany-Witten transitions \cite{HananyWitten}.  This is not to say
that the two types of branes have physically different properties, but
rather that the K-theory formalism treats them differently. In particular, if our spacetime $M$ is compact 
and has no boundary, as it will be during most of this paper, then because $F_p$ is gauge-invariant Stokes 
theorem tells us that
\beq
\int_M dF_p=\int_{\partial M} F_p=0\,.
\eeq
Therefore in the compact case the right hand side of Eq.~(\ref{meglio}) vanishes only when the
D-brane charge vanishes, at least up to the torsion terms that we have been neglecting.  This leads us to the claim
\newline

\noindent
\claim {\textit{On compact spacetimes fluxes can be classified by
twisted K-theory only if the total D-brane charge is torsion.}}
\newline

\smallskip
\noindent
We will see examples in which compact spacetimes may have torsion D-brane charge which is nonvanishing after the torsion corrections that we will describe momentarily.  Twisted K-theory will not classify fluxes in
these cases.  
%We note that we cannot exclude 
%the possible existence of
%gravitational contributions to 
%the brane charges in Eq.~(\ref{meglio}).  

While twisted K-cohomology classifies fluxes, twisted K-homology
classifies the branes that source these fluxes.  That is to say, it
classifies the $dF$-type branes.  The $H\wedge F_p$ type branes are
quotiented away when passing from homology to K-homology because these
branes are created by changing $F_p$ and the K-classification of
branes applies only when we forget about globally defined field strengths $F_p$, just as the K-classification of fluxes required that we forget about globally defined connections $C_{p-1}$.

The analogous condition to (\ref{speciale}) in the case of branes is obtained by replacing the flux $G_p$ with the charge $dG_p$.  It is just the condition that the Freed-Witten anomaly \cite{FW} vanishes, or equivalently the condition that the brane is not a baryon.  We will use the word baryon \cite{baryons} to mean a brane on which other branes end, where the terminology came from the fact that they correspond to baryonic vortices in the worldvolume gauge theories of some probe branes.  

Physical D-branes must satisfy yet another condition, which is not in
general satisfied by twisted K-theory classes, that the flux $F$ that they source is globally defined.  In particular this means that there is no net $dF$-type D-brane charge on a compact spacetime, and so the branes are not classified by twisted K-theory.  However the twisted K-theory of a compact spacetime is still instructive for two reasons.  First, while there is no net charge there may still be D-branes.  We will see that some properties of D-branes, such as the lower brane charges that they carry and the possible remnants when they annihilate, are described by (the extension problem of) twisted K-theory even when the spacetime is compact. Secondly, often there is a similar configuration that is noncompact, for example in the cases $k=1$ and $k=2$ of the present paper there is an extra sphere that plays no role and may be replaced by a noncompact manifold, in which case the twisted K-homology classes described may yield honest D-branes.

\subsection{Powers of $H$ and the S-duality Puzzle}

The RR gauge transformations (\ref{trasformazioni}) that lead to the twisted K-theory classification of RR fluxes and D-branes are not the only gauge transformations available in type II supergravities.  For example the type IIB action is invariant under $SL(2,\R)$ S-duality transformations that mix the RR 3 form $G_3$ and the NS 3-form $H$ and the S-duals of the gauge transformations (\ref{trasformazioni}) yield distinct gauge transformations.  It has been conjectured \cite{HT,H,Sch} that an $SL(2,\Z)$ subgroup is a symmetry of the full quantum theory.  If this is true then one may wish to define a twisted K-theory corresponding to the image of Eq.~(\ref{trasformazioni}) under each element of $SL(2,\Z)$.  Such dual K-classifications have been applied to stacks of D3-branes on an orientifold 3-plane in Ref.~\cite{Oscar} and to the Klebanov-Strassler geometry \cite{Kleb} in Ref.~\cite{EvsMay}.  While the $\{T^n\}=\Z\subset SL(2,\Z)$ subgroup
\beq
G_3\mapsto G_3+n H\hsp H\mapsto H\,,
\eeq
acts trivially on the gauge transformations and the K-theory classes, the orbit of the $SL(2,\Z)$ actions on the gauge transformations still yields an infinite number of different twisted K-theories all of which may simultaneously classify the flux and brane spectra of a given spacetime.  That is to say, a given configuration of fluxes of branes may, if it satisfies the criteria given in the last subsection, correspond to an element of each member of an infinite family of twisted K-theories \cite{EvsFlux}.  We will refer to this family as S-covariant K-theory. 

Such a configuration must satisfy an infinite number of conditions that are S-dual to $(\ref{speciale})$.  For example
\beq
(Sq^3+G_3)H=0\hsp (Sq^3+G_3)G_5=0\hsp (Sq^3+G_3+H) G_3= H\cup G_3=0\,. \label{condizioni}
\eeq
Note that the $\Z\subset SL(2,\Z)$ subgroup that acts trivially on the gauge transformation also acts trivially on these constraints.  Combining the first and last conditions of Eq.~(\ref{condizioni}) with that of Eq.~(\ref{speciale}) we find in particular that
\beq
G_3\cup G_3=G_3\cup H=H\cup H=0\, \label{quadratico}
\eeq
for all fluxes in S-covariant K-theory.

Eq.~(\ref{quadratico}) is stronger than the condition that D3-brane charge
vanishes.  To find this condition, we begin with a compact manifold
with $H=G_3=0$.  For simplicity we assume that the fundamental group
is pure torsion so that the first cohomology group $\textup{H}^1$ is trivial 
and so the field $G_1$ is trivial and
the dilaton is globally defined.  The manifold is compact and so there
are no $dG_5$ type D3-branes, and the vanishing $G_3$ and $H$ insure
that there are no Hanany-Witten type D3-branes.  We cannot exclude the
possibility that there is brane charge that results entirely from
gravitational effects, which would be determined by the topology of
the spacetime.  For example Green-Schwarz-like terms may correct the equations of motion (\ref{meglio}), such as the one-loop term in the IIA supergravity action calculated in Ref.~\cite{VW} which contributes to the fundamental string charge \cite{SVW}.  Note that simply dualizing that contribution to obtain gravitational D-brane charge fails as fluxes are inevitably produced by the duality, but an implicit formula obeyed by the gravitational charge appears in Ref.~\cite{DFM}.  We let $P$ denote the 6-form dual to the gravitational D3-brane charge.

\begin{figure}[ht]
\begin{center}
\leavevmode
\epsfxsize 14   cm
\epsffile{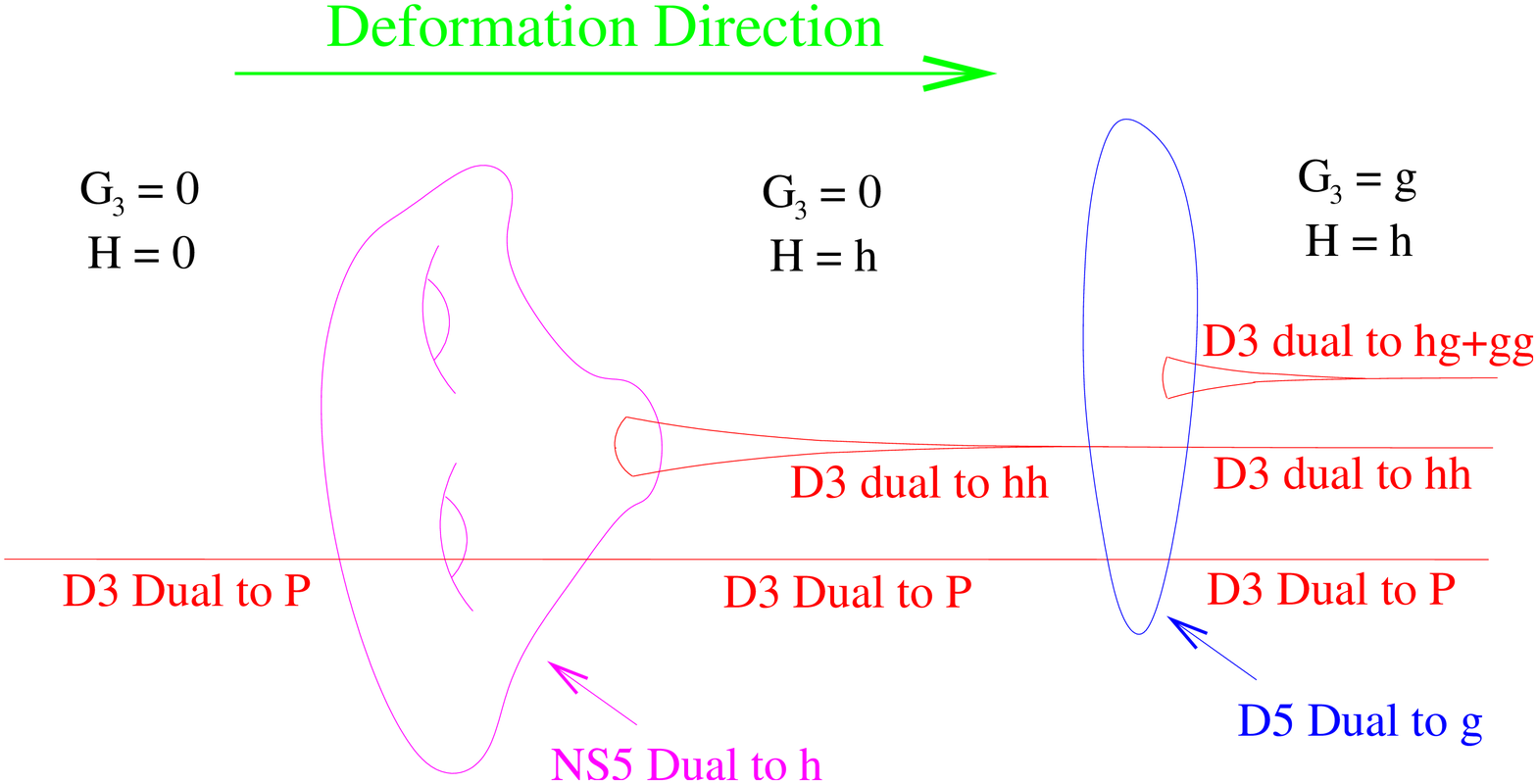}    
\end{center} 
\footnotesize{\caption{\footnotesize{On the left the fluxes are turned off and the D3-brane charge $P$ is determined by the topology of the spacetime.  As we move to the right, an NS5 sweeps out the cycle dual to the 3-class $h$, and so to the right of the NS5 there is an NS flux $H=h$.  The Freed-Witten anomaly of the NS5-brane is canceled by inserting a D3 dual to $hh$, and so the D3 charge changes by $hh$.  Next we pass a D5-brane dual to $g$, whose Freed-Witten anomaly is $hg+gg$.  The anomaly requires that the D3 charge jump by $hg+gg$.  At every stage the D3 charge is $P+HH+G_3H+G_3G_3$, and on the right $G_3$ and $H$ are arbitrary.}}}
%\caption{This is another view of Figure.}
\end{figure}

Now we may turn on any $H$ and $G_3$-flux by letting 5-branes sweep
out the dual cycles.  First we turn on the $H$-flux with an
NS5-brane.  The NS5-brane in IIB carries a $U(1)$ gauge field under
which fermions are charged that are in the $spin$ lift of the normal
bundle.  If $H\cup H$ is nonvanishing then this $spin$ lift does not
exist and furthermore no shift of the $U(1)$ gauge field can render the partition
function well-defined.  This is the Freed-Witten anomaly \cite{FW} and
it can be canceled by including a D3-brane dual to $H\cup H$ which
ends on the NS5.  Thus the NS5 sweeps out a cycle and vanishes,
leaving behind not only the desired $H$-flux, but also $H\cup H$ units
of D3 charge.  Now we may turn on any $G_3$-flux by sweeping a
D5-brane through a dual cycle.  The D5-brane has a Freed-Witten
anomaly $G_3\cup G_3$ resulting from its possible failure to be
$spin^c$ and a further anomaly $G_3\cup H$ resulting from the $H$ flux
on its compact worldvolume.  Again these anomalies are canceled by D3-brane
insertions.  Thus we are able to create an arbitrary $G_3$ and $H$-flux, but in the process we automatically change the D3-brane charge.

\noindent
\claim {\textit{Type IIB on a compact spacetime with torsion
fundamental group has D3 charge equal to $H\cup H+H\cup G_3+G_3\cup G_3+P$.}}\\

\noindent

This claim extends a result proven in Ref.~\cite{DFM} to more general backgrounds, in particular to backgrounds with D-branes and to backgrounds not dual to M-theory compactified on a 2-torus.  Note that while $H\cup G_3$ is the familiar term from tadpole cancellation in the flux compactifications literature, this claim suggests that the other 3 terms appear as torsion corrections to the usual formula.  
This would 
kill most flux vacua. However, it is possible that it would not change 
scales by enough orders of magnitude to be detected by landscape considerations.

If the D3-brane itself wraps nontrivial $G_3$ or $H$-flux then it also
suffers from a Freed-Witten anomaly which must be canceled by
inserting F-strings or D-strings respectively.  The string worldsheets are tubes with one end wrapping the dual of the offending flux in the D3 worldvolume.  The spacetime is compact and so the string must have two ends.  The other end of the string must be on a second brane, where it will act as some kind of source in the worldvolume gauge theory.  This second brane is compact, as the spacetime is compact, and so the charge needs to be canceled by a worldvolume flux on the second brane.  The compactness of spacetime also means that the second brane cannot be a source brane, but rather is a Hanany-Witten type brane.

If for example the string is an F-string then the other end may be on a second brane which is a D1 that wraps a cycle with $G_1$ flux or else a D5 that wraps a cycle with a five-form flux.  The second brane is not a source brane, and so in particular the D5 must satisfy
\beq
Q_{{\textup{\footnotesize{D}}}5}=H\cup G_1\,.
\eeq
In either case we see that $G_1$ needs to be nontrivial for this
cancellation to occur, and so the spacetime must have $\H^1\neq0$ and so
the fundamental group must not be pure torsion.  Thus if the fundamental group is torsion then there is no candidate for the second brane and so the anomaly cannot be canceled.  Therefore the D3-charge
cupped with $G_3$ and $H$ must vanish, which leads to our final claim\\

\noindent
\claim {\textit{Type IIB on a compact spacetime with torsion
fundamental group is consistent only if $G_3\cup G_3\cup G_3+G_3\cup
P$ and $H\cup H\cup H+H\cup P$ vanish.}}\\

\noindent

In this paper we will test these claims by considering three examples of type IIB string theory 
backgrounds, 
$\rpt\times\rp^{2k+1}\times S^{6-2k}$ with torsion $H$ flux, which have $H\neq 0$, $H\cup H\neq 0$ and $H\cup H\cup H\neq 0$ for $k=1,\ 2$ and $3$ respectively.  These backgrounds are particularly simple to understand because \cite{MW}, as we will show in Sec.~\ref{t-dualita}, they are T-dual to $S^3\times S^{2k+1}\times S^{6-2k}$ with no fluxes at all.  The example $k=2$ is not $spin$, and so the definition of string theory on this space is not obvious, although one might guess that the $H$ flux provides some kind of generalized $spin^c$ structure.  This is the case after dimensionally reducing on a two-torus, the two 2-forms resulting from the dimensional reduction of $H$ are the curvatures of the two dual circle bundles whose total space is $spin$.  As in Ref.~\cite{DLP}, the T-duality to $S^3\times S^5\times S^2$ with no flux may imply that the string theory is well defined and even supersymmetric via ``supersymmetry without supersymmetry''.   However \cite{MW}, as in \cite{DLP}, the dual spheres are smaller than the string scale and so it is not clear that they provide a definition.  Instead it may be necessary to define this background as an orientifold.  In this case some fields will be valued in $\Z_2$ twisted cohomology \cite{baryons,HK}. 

We will describe the D3-brane insertions in Sec.~\ref{inserzioni} and then in Sec.~\ref{inserzioniduale} we will use the T-duality prescription of Refs.~\cite{BEMa,BEMb} to follow these and other D-branes through the T-dualities.  This allows us to compare the above conjectures about Freed-Witten anomalies on projective spaces to the better understood physics of the compactification on a product of spheres, where the absence of torsion and of fluxes means that none of these anomalies are present.

A central role is played by the twisted K-theory classification of branes.  In Sec.~\ref{k-teorie} we will use the Atiyah-Hirzebruch spectral sequence to compute the twisted K-theories of each space and demonstrate that as expected T-duality shifts the dimension by one.  This map on twisted K-theory will then allow us to confirm that we have correctly T-dualized each of the branes.  Many of these computations are new and we hope that they will be of independent interest.

Along the way we will run into a number of interesting phenomena.  For example in Sec.~\ref{riscaldimento} we will, following the suggestion of Ref.~\cite{Bergman}, identify the lower brane charges carried by D-branes with the solution to the K-theory extension problem.  In particular we will see an example in which a lower half-brane charge is carried when the normal bundle is $spin$ and the $B$-field is trivial, which is possible because of a factor of two that appears in the spectral sequence.  In Sec.~\ref{t-dualita} we will find an example in which the differential $d_5$ acts nontrivially, and we will see that its action agrees with the conjectured form of $d_5$ in Ref.~\cite{Uday}.  We will also find, in Sec.~\ref{inserzioniduale}, a composite Freed-Witten anomaly that creates a topologically nontrivial charge from a topologically trivial flux (although the deformation to the trivial flux does not respect a free circle action) and a gauge-dependent flux.  We will use this anomaly to partially extend the 
Diaconescu-Moore-Witten (DMW) \cite{DMW} anomaly $W_7=0$ to the non-$spin$ case.  In the appendix we summarize the relevant properties of projective spaces.

\begin{figure}[ht]
\begin{center}
\leavevmode
\epsfxsize 14   cm
\epsffile{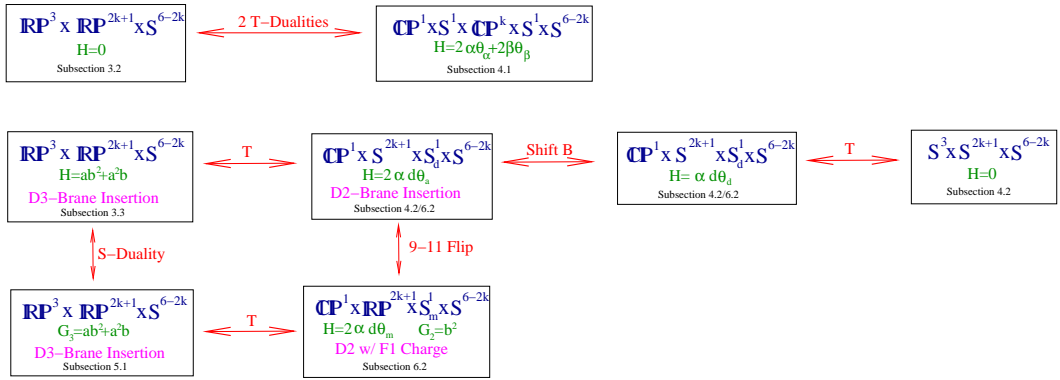}    
\end{center} 
\footnotesize{\caption{\footnotesize{This is a summary of the main compactifications that appear
in this note and the dualities and transformations that relate them.}}}
%\caption{This is another view of Figure.}
%\label{2dmu} 
\end{figure}

% =========================================================================
\section{Warm up: Type IIB on $\rps\times S^3$ Without Flux} \label{riscaldimento}

In this section we will classify RR fluxes and branes in type IIB string theory on $\rps\times S^3$ with no NS flux.  The relation between branes of various dimensions that will be crucial in the main argument of this note may already be seen in this simpler example.  We begin by using the Atiyah-Hirzebruch spectral sequence (AHSS) and K\"unneth theorem to review the relevant untwisted K-theories.  Then we interpret these groups physically in terms of fluxes and also in terms of branes carrying the charges of lower-dimensional branes.

% =========================================================================
\subsection{Calculating Untwisted K-groups}

In Ref.~\cite{AH} Atiyah and Hirzebruch showed that the associated graded 
K-theory of a space 
$X$, $Gr(K(X))$, which is related to K-theory by an extension problem, may be calculated from the cohomology of $X$ by taking a series of subsets and quotients.  In this example, unlike the examples that we will encounter later, this process is trivial and $Gr(K(X))$ is in fact just the original integral cohomology, or more precisely the sum of either the even or the odd cohomology. The relevant homology and cohomology groups are  
\bea
&&\H_0(S^3)=\H_3(S^3)=\H_0(\rps)=\H_7(\rps)=\Z \,, \nonumber\\
&&\H_1(\rps)=\H_3(\rps)=\H_5(\rps)=\Z_2\,,\nonumber\\
&&\H^0(S^3)=\H^3(S^3)=\H^0(\rps)=\H^7(\rps)=\Z\,, \nonumber\\
 &&\H^2(\rps)=\H^4(\rps)=\H^6(\rps)=\Z_2 \,,
\eea
and so the associated graded K-groups are
\bea
&&Gr(K_0(S^3))=Gr(K_1(S^3))=Gr(K^0(S^3))=Gr(K^1(S^3))=\Z\,,\nonumber\\
&&Gr(K_0(\rps))=\Z\hsp Gr(K_1(\rps))=\Z\oplus\Z_2^3\,,\nonumber\\
&&Gr(K^0(\rps))=\Z\oplus\Z_2^3\hsp Gr(K^1(\rps))=\Z\,.
\eea
To obtain the actual K-groups we need to solve a series of extension
problems.  The ones that will be nontrivial are described by the short
exact sequences
\bea
&&\H_1(\rps)=\Z_2\rightarrow F_2\rightarrow H_3(\rps)=\Z_2\hsp
F_2\rightarrow F_3\rightarrow H_5(\rps)=\Z_2\,,\nonumber\\
&&\H^6(\rps)=\Z_2\rightarrow F^2\rightarrow H^4(\rps)=\Z_2\hsp
F^2\rightarrow F^1 \rightarrow H^2(\rps)=\Z_2\,.
\eea 
The solutions to these extension problems are not unique, and so we will 
state the answers which are derived in, for example, Ref.~\cite{Bra} and 
then explain the underlying physics. The solution is 
\beq
F_2=F^2=\Z_4\hsp F_3=F^1=\Z_8 \,,
\eeq
and so the three $\Z_2$ torsion groups in each $Gr(K)$ assemble into a single $\Z_8$ in the actual K-groups
\bea
&&K_0(S^3)=K_1(S^3)=K^0(S^3)=K^1(S^3)=\Z\,,\nonumber\\
&&K_0(\rps)=\Z\hsp K_1(\rps)=\Z\oplus\Z_8\,,\nonumber\\
&&K^0(\rps)=\Z\oplus\Z_8\hsp K^1(\rps)=\Z\,.
\eea
This means that if we interpret the generator $x$ of $\H^2(\rps)=\Z_2$ as the generator of the $\Z_8\subset K^0(\rps)$ then in K-theory instead of being order 2, $x$ is order 8.  Thus while $2x\in \H^2(\rps)=\Z_2$ is the trivial element in cohomology, as an element of K-theory it is nontrivial and corresponds to the generator $y$ of $\H^4(\rps)=\Z_2$, while $3x$ in K-theory corresponds to $x+y$ in cohomology.  

So far this is just what one would find by associating the elements of $\H^{2k}$ with the Chern classes $c_k$ of the corresponding K-class and defining addition to be the direct sum of the corresponding bundles.  However this naive association fails for $4x$, which corresponds to the generator $z$ of $\H^6(\rps)=\Z_2$ in the spectral sequence.  This cannot be the third Chern class because \footnote{We will denote by $sq^k$ the Steenrod square in 
cohomology with $\Z_2$ coefficients, and by $Sq^{2k+1}$ the lift of $sq^{2k+1}$ to
integer coefficients.}
\beq
c_3=c_1\cup c_2+sq^2 c_2\ \mod 2\,,
\eeq
and so when $c_1=c_2=0$ the third class $c_3$ must be even and thus
trivial in $\H^6(\rps)=\Z_2$.  We will argue that the class $z$
corresponds not to $c_3$ but to $ch_3=c_3/2$ which is odd. Ordinarily,
the Chern characters are only elements of rational cohomology and thus
do not see torsion classes, which would mean that $ch_3$ ordinarily
would miss the $\Z_2$.  However, we claim that in the cases in which
Chern characters compute charges in string theory they do admit a lift
to integral cohomology (with normalization defined by the AHSS) and so are sensitive to
torsion.  Here this is a result of the evenness of $c_3$, but in
general it will reflect that fact that they may be expressed as the
images of products of Steenrod squares and Steenrod powers.  

Similarly, the $\Z_8$ subgroup of $K_1(\rps)$ is generated by the 
generator $x$ of $\H_5(\rps)=\Z_2$, which is order 2 in homology and order 8 in K-homology.  $2x$ generates $\H_3(\rps)=\Z_2$ while $4x$ generates $\H_1(\rps)=\Z_2$, where again we will see a crucial factor of two.

We may now combine the K-theories of the sphere and projective space to find the K-theory of the product using the K\"unneth formula.  The result is just the tensor product of the original K-theories because the K-theory of the sphere has no torsion and so the $Tor$ term in the K\"unneth formula is trivial
\bea
&&K_0(\rps\times S^3)\pic=\pic \big(K_0(\rps)\otimes K_0(S^3)\big) \oplus \big(K_1(\rps)\otimes K_1(S^3) \big) \pic=\pic\Z^2\oplus\Z_8 \,, \nonumber\\
&&K_1(\rps\times S^3)\pic=\pic\big(K_0(\rps)\otimes K_1(S^3)\big) \oplus \big(K_1(\rps)\otimes K_0(S^3) \big)\pic=\pic\Z^2\oplus\Z_8 \,, \nonumber\\
&&K^0(\rps\times S^3)\pic=\pic\big(K^0(\rps)\otimes K^0(S^3)\big)\oplus \big(K^1(\rps)\otimes 
K^1(S^3) \big)\pic=\pic\Z^2\oplus\Z_8 \,, \nonumber\\
&&K^1(\rps\times S^3)\pic=\pic\big(K^0(\rps)\otimes K^1(S^3)\big)\oplus \big(K^1(\rps)\otimes K^0(S^3) \big)\pic=\pic\Z^2\oplus\Z_8\,.
\eea
The extension problem now means that, for example, the generator
\beq
x\in \H^5(\rps\times S^3)=\H^2(\rps)\otimes\H^3(S^3)=\Z_2 \,,
\eeq
is order two in cohomology but order eight in K-theory.  So for example $2x$ is the generator of $\H^4(\rps)\otimes\H^3(S^3)$.

The untwisted K-theories of other $\rp^{2k+1}$'s are obtained identically, except that there are $k$ factors of $\Z_2$ and so $\Z_8$ is replaced by $\Z_{2^k}$.

% =========================================================================
\subsection{Branes and Fluxes on $\rps\times S^3$.}

In type IIA string theory D-branes are classified by $K_1$ while RR field 
strengths are classified by $K^0$, both of which in our case are 
$\Z^2\oplus\Z_8$.  Thus all branes and fluxes are generated by three 
elementary ones, the generators of the respective K-groups, one of which 
is order eight.  For example the fluxes are generated by $G_0$, which is 
the Romans' mass \cite{Rom}, $G_{10}$, which again is the Romans' mass
(in the dual sense), 
and 
finally $G_2\in \H^2(\rps)\otimes \H^0(S^3)$ which is torsion.  The Romans' mass may appear twice because K-theory does not really classify fluxes in string theory, which cannot all be simultaneously quantized, but rather on a toy model of string theory in which the self-duality condition
\beq
G_p=*G_{10-p}
\eeq
is not imposed.  This omission is crucial as the Hodge star varies continuously with the metric and so is incompatible with the simultaneous quantization of the Chern characters that is inherent in the K-theory classification.  In the physical string  theory \cite{Wita} only half of the fluxes may be quantized at a time, but on a compact space one needs to check that the partition function is invariant under the choice of which half is quantized, as in the case in the chiral scalar 2-dimensional field theory.  Thus one should expect K-theory to double-count the simultaneously observable degrees of freedom.

$G_2$ is the Chern class of the M-theory circle which is fibered over
 $\rps\times S^3$.  While there are only two possible fibrations of
 the M-theory circle, corresponding to the two elements of $\H^2$, as
 an element of K-theory if we add together the two nontrivial M-theory
 bundles we find that while the M-theory bundle becomes trivial, we
 now have a unit of $G_4$-flux.  Similarly adding together two units
 of $G_4$-flux we find a unit of $G_6$. For a global version of the 
reduction of M-theory to type IIA leading to twisted K-theory, see 
\cite{MS}.

Although adding fluxes may seem a bit abstract if not ill-defined, we may find the same result by physically adding two D-branes that source the corresponding fluxes.  The D-branes are classified by $K_1=\Z^2\oplus\Z_8$ and so again there are three generators.  There is a D2-brane that wraps the $S^3$, a D6-brane that wraps the $\rps$, and finally there is a torsion D4-brane that wraps the submanifold $\rp^5\subset\rps$.  While a homology classification of D-branes would tell us that by deforming the configuration two D4-branes can annihilate to nothing, as they are each $\Z_2$ charged, in fact after the annihilation a D2-brane remains that wraps $\rpt\subset\rps$.  Similarly two D2-branes may annihilate to leave a D0-brane wrapping the circle $\rp^1$.  However if two D0-branes annihilate, nothing remains.  Thus we say that the D4-branes each carry half of a unit of D2-brane charge while the D2-branes each carry half a unit of D0-brane charge.  Note that while the element 1 corresponds to a D4-brane with half a unit of D2 charge, the unit 7 corresponds to a D4-brane with negative half a unit of D2-charge, and so the two D4-branes corresponding to 1 and 7 may annihilate to a state that contains no D2-branes.  However there is no such confusion if we choose orientation conventions (the SUSYs that will be preserved by branes) and write all states in terms of branes and not antibranes, so that 7 is written as a D4-brane plus a D2-brane plus a D0.

This explains the meaning of two units of $G_2$-flux combining to form a 
unit of $G_4$-flux.  We may consider a D6-brane wrapping $\rp^5\times S^3$ that sources the $G_2$ 
flux.  There is no space in $\rps\times S^3$ for such a D6-brane and in fact it has one too many dimensions, however we are free to add another dimension $v$ that is parameterized by 
deformations \cite{Witb, Hor}.  In fact, one such choice of dimension that 
appears naturally in applications is the Renormalization Group (RG) flow \cite{EvsMay}. That is 
to say that a $G_2$-flux measured at one energy scale (a dimensional 
reduction of M-theory defined by some characteristic distance) may differ 
from the $G_2$-flux measured at another scale, and this difference 
corresponds to a dynamical process\footnote{Here ``time'' refers to the RG 
direction, so for example the process may occur as the system relaxes.  
More generally K-classes appear to correspond to universality classes in 
the worldvolume theories.} in which a D6-brane has nucleated, swept out a linking 8-cycle, and collapsed out of existence.  In the Klebanov-Strassler cascade \cite{Kleb}, for example, the relation between the brane position and energy scale is a consequence of the RG flow of the coupling constant of the worldvolume effective gauge theory.  

Thus the 10-dimensional cross-section at one value of $v$ enjoys a nontrivial $G_2$-flux generating $H^2(\rps)$, but as one moves in the $v$ direction one passes a D6-brane (corresponding to $1\in\Z_8$), wrapping $\rp^5\times S^3$, and the 
$G_2$-flux turns off.  One may pass a second D6-brane and the flux turns back on.  If this second D6-brane corresponds to the element $1\subset\Z_8$, and not the element $7\subset\Z_8$ then the sum of these two D6-branes carries a unit of D4-brane charge.  This means that as we pass both of these branes $G_4$ jumps by a unit.  This is the meaning of the fact that two units of $G_2$-flux add to a unit of $G_4$-flux, and also the source of the caveat that we need to be sure that our nontrivial $G_2$'s both lift to the same element in K-theory.

% =========================================================================
\subsection{Solving the Extension Problem with Physics}

We have argued that the physical interpretation of the nontrivial solution of the extension problem is that the D4-brane wrapping $\rp^5\subset\rps$ carries half a unit of D2 charge, while the D2 wrapping $\rpt$ carries half a unit of D0 charge.  It should be possible to understand these charges, and thus the solution to the extension problem, in terms of the worldvolume physics of the corresponding branes.  For example, finding the D2 charge in the D4-brane is routine.  The D4-brane worldvolume theory contains fermions that are charged under the worldvolume U(1) gauge field and also are valued in the spinor representation of the normal bundle.  The $S^3$ part of the normal bundle is trivial, however the normal bundle $N$ of $\rp^5\subset\rp^7$ is not $spin$.  This means that there is an obstruction to lifting the normal bundle to a $spin$ bundle, and so the consistency of the fermion partition function demands that the $U(1)$ bundle enjoy an obstruction that cancels that of the normal bundle.  That is to say, the quantization of the $U(1)$ gauge field is shifted
\beq
F\in\Z+\textstyle{\frac{1}{2}}\,.
\eeq
We may now use the D4-brane worldvolume coupling
\beq
S\supset\int_{\rp^5} F\cup C_3=\frac{1}{2}\int_{\rpt}C_3\,,
\eeq
to conclude that the D4-brane carries half a unit of D2 charge.

In general the obstruction to being $spin$ is the second Stiefel-Whitney class $w_2$, and so $F$ is shifted by $w_2(N)$, yielding $w_2(N)$ units of D$p$-brane charge inside of a D$(p+2)$-brane, although we will soon see that the normalization is very subtle.  We may calculate this obstruction directly from the cohomology of $\rps$.  If we say that $a^2$ is the generator of $\H^2(\rps)=\Z_2$, so that $a^4$ and $a^6$ generate $\H^4$ and $\H^6$, respectively, then the D4-brane is the Poincar\'e dual (PD) of $a^2$.  To calculate the D2-brane charge corresponding to a D4-brane PD($a^2$) we need to calculate $w_2$ of the normal bundle of PD($a^2$), which gives the worldvolume field strength, and then push that forward onto the whole space so that we get the divisor of the corresponding $U(1)$ gauge bundle (the codimension two subset of our D4 that corresponds to the $F$-flux), which is our potential D2-brane.  This operation is done by the cohomology operation $sq^2$, and so we may argue that the D2-charge is given by
\beq
{\text{PD}}(\textup{D}2)\stackrel{?}{=}sq^2 a^2=a^4\,.
\eeq
This is nonzero and so we see the D2-brane charge, however we have not been careful about the normalization.  In fact the extension problem told us that the D4 should carry only half of a D2-brane.

This argument applied to the D0-brane charge in the D2-brane fails.  The problem is that the D2-brane wraps $\rpt\subset\rps$ whose normal bundle is the sum of the normal bundles of $\rpt\subset\rp^5$ and $\rp^5\subset\rps$.  While neither summand is $spin$, the $w_2$'s of the summands cancel and so the normal bundle of the D2-brane is $spin$. It then seems as though the D2-brane should carry integral D0-charge.  In terms of the $sq^2$ construction we see the same unfortunate cancellation
\beq
{\text{PD}}(\textup{D}0)\stackrel{?}{=}sq^2 sq^2 a^2=sq^2 a^4=(sq^2 a^2)a^2+a^2(sq^2 a^2)=a^6+a^6=2a^6=0 \,, \label{doppio}
\eeq
where the last equality comes from the fact that $a^6$ generates the
$\Z_2$ torsion group $\H^6(\rps)$.  The two $a^6$'s that cancel are
the two $w_2$'s of the normal bundle summands.  However the overall
normalization of the worldvolume D$p$ charge from the Wess-Zumino
terms of the worldvolume action is a Chern character and so lives in
rational cohomology and has no natural normalization\footnote{It may
be interesting to see how the inclusion of the $\sqrt{\hat{A}}$ term
changes this situation.} until we attempt to apply Dirac quantization.  That is, the lift of the charge to an integral class and so the normalization of the charge should be derivable from the physics, and vice versa.

We know from the solution to the extension problem that the correct normalization yields the nontrivial D0 charge $a^6$ inside of our D2-brane or D4-brane.  This is one half of the value found in Eq.~(\ref{doppio}).  To see that this half is canonically defined we will need to include it before we make the transition from integral to rational cohomology, that is in the Steenrod square expression
\beq
{\text{PD}}(\textup{D}0)=\textstyle{\frac{1}{2}}\, sq^2\, sq^2\, a^2\,.
\eeq
This is the same factor of 1/2 that appeared in the formula for $d_5$
in Ref.~\cite{Uday}.  As in that case, the division by two may be
defined by rewriting the expression in terms of the
Steenrod\footnote{Acting on 3-torsion $P^1$ is the usual cube, which
yields the Milnor primitive term $Q_1=\beta P^1$ in $d_5.$  However
acting on 2-torsion we define it to be a secondary operation that only acts
on the cohomology of $d_3$.}  cube $P^1$, which cubes 2-cocyles modulo 3
\beq
{\text{PD}}(\textup{D}0)=P^1 {\text{PD}}(\textup{D}4)=P^1 a^2 = a^6\,. \label{powerfrac}
\eeq

This agrees with the interpretation of branes that are not closed under the AHSS differentials $d_{2k+1}$ as baryons.  For example if $b$ is Poincar\'e dual to a D$p$-brane and
\beq
d_3 b=Sq^3\, b=\beta\, sq^2\, b =c \neq 0 \,,
\eeq
then the D$p$-brane carries the charge of a half D$(p-2)$-brane wrapped about PD($sq^2\, b$).  However PD($sq^2\, b$) does not lift to integral homology because it is not in the kernel of $\beta$ and so the half D$(p-2)$-brane, being oriented, has a boundary $(PD(c))$ inside the worldvolume.  This means that the half-brane must continue past this boundary into the bulk (wrapping PD($c$)) and so the D$p$-brane is a baryon.  

The extra factor of 2 allows us to extend this story to $d_5$ if
\beq
d_5 b=\beta P^1 b =c \neq 0.
\eeq
Generalizing Eq.~(\ref{powerfrac}) the D$p$-brane dual to $b$ carries $P^1b$ 
units of D$(p-4)$ charge.  The units are smaller by a factor of two than those obtained by iterating the $d_3$ argument twice, but even this quarter D$(p-4)$-brane has no boundary when $c\neq 0$ and so continues into the bulk, creating a baryon.  The fact that D$p$-branes not closed under $d_5$ yield D$p$-D$(p-4)$ baryons was first noted in Ref.~\cite{MMS}.  It should be possible to extend this reasoning to all of the differentials, such as the yet higher Steenrod powers that are relevant for string theory on $\rp^9\times S^1$.  Including a $B$-field in this argument may lead to the construction of the twisted differentials as the Bockstein of the corresponding lower brane charges, which may be decomposed into $B$ terms and the lower-dimensional brane charges of the untwisted case following the strategy of Ref.~\cite{AtiyahSegal}.

% =========================================================================
\section{Computing the Twisted K-theories of $\rpt\times\rps$}  \label{k-teorie}

We are ultimately interested in the twisted K-theories of $\rpt\times\rp^{2k+1}\times S^{6-2k}$ 
($k=1,2,3$).  However, all of these may be obtained from the those of $\rpt\times\rps$ by removing $3-k$ of the $\Z_2$'s and using the K\"unneth formula to include the extra sphere.  Thus in this section we will only compute the twisted K-theory of $\rpt\times\rps$ and we will state the results for the other cases.

% =========================================================================
\subsection{Computing the Cohomology of $\rpt\times\rps$}

The real projective space $\rpn$ is the quotient of the $n$-sphere $S^n$ by the antipodal map.  The cohomology ring with $\Z_2$ coefficients is just a polynomial ring over $\Z_2$
\beq
\H^*(\rpn;\Z_2)=\Z_2[a]/(a^{n+1}) \,, \label{poly}
\eeq
where $a$ is the 1-dimensional cocycle.  Intuitively, $a^k$ is supported on the $\rp^k\subset\rpn$.   The $\Z$ homology and cohomology are only slightly more complicated.  Again they are generated by the $\rp^k$ subsets, but with the boundary map
\beq
\partial:C_{2k}\rightarrow C_{2k-1}:a_{2k}\mapsto 2a_{2k-1} \,,
\eeq
acting on the generator $a_{2k}$ of the $2k$-chains, intuitively the subset $\rp^{2k}$.  Dualizing, one obtains the coboundary map on the $(2k-1)$-cochains
\beq
\partial^\dagger:C^{2k-1}\rightarrow C^{2k}:a^{2k-1}\mapsto 2a^{2k}\,.
\eeq
Thus the even cycles and odd cocycles are zero, while the odd cycles and even cocycles are boundaries and coboundaries if they are multiples of two.  If $n$ is odd, as it will be in this example, then the top (and as always the bottom) dimensional chain and cochain are in the kernel and never the image of the differential, and so the top dimensional homology and cohomology is $\Z$.  

We have then found, for the cases $n=3$ and $n=7$, that the nontrivial classes are
\bea
&&\H_0(\rpt)=\H_0(\rps)=\H_3(\rpt)=\H_7(\rps)=\Z \,, \nonumber\\
&&\H_1(\rpt)=\H_1(\rps)=\H_3(\rps)=\H_5(\rps)=\Z_2 \,, \nonumber\\
&&\H^0(\rpt)=\H^0(\rps)=\H^3(\rpt)=\H^7(\rps)=\Z \,, \nonumber\\
 &&\H^2(\rpt)=\H^2(\rps)=\H^4(\rps)=\H^6(\rps)=\Z_2\,.
\eea
Our $\Z_2$ cohomology ring generator $a$ is not a cocycle now that we are using $\Z$ coefficients because it has coboundary $\partial^\dagger a=2a^2\neq 0$.  However we may still evaluate each term in the cohomology ring modulo two and so identify it with a subring of the $\Z_2$ cohomology.  This allows us to write the $\Z$ cohomology generator of $\H^k$ in terms of the $\Z_2$ generator $a$, basically $\H^k$ is generated by $a^k$.  Since we know how to multiply the $\Z_2$ generators (\ref{poly}) we can then guess how to multiply the $\Z$ generators.  

We let $a$ and $b$ be the generators of $\H^1(\rpt,\Z_2)$ and $\H^1(\rps,\Z_2)$, respectively.  Although $a$ and $b$ themselves do not lift to integral classes, we will write the generators of the integral classes as powers of $a$ and $b$.  Then we may encode the ring structure in the following definition, where $\gnr{x}$ is the additive group generated by $x$
\bea
\H^0(\rpt)=\gnr{1}\hsp& \H^2(\rpt)=\gnr{a^2}\hsp& \H^3(\rpt)=\gnr{a^3}\,, \\
\H^0(\rps)=\gnr{1},\ \H^2(\rps)=\gnr{b^2},&\hspace{-.5cm} \H^4(\rps)=\gnr{b^4},&\hspace{-.5cm} \H^6(\rps)=\gnr{b^6},\  \H^7(\rps)=\gnr{b^7}\nonumber\,. \label{rpgen}
\eea

We ultimately want the cohomology of the product.  To find this, we will first evaluate the homology of the product using the K\"unneth formula
\beq
\H_n(\rpt\times\rps)=\bigoplus_i \big( \H_i(\rpt)\otimes \H_{n-i}(\rps) \big) \bigoplus 
\big( \bigoplus_i Tor(\H_i(\rpt),H_{n-i-1}(\rps)) \big) \,,
\eeq
where the only nontrivial $Tor$ term will be
\beq
Tor(\Z_2,\Z_2)=\Z_2.
\eeq
The $Tor$ terms contribute a $\Z_2$ to $\H_3$, $\H_5$ and $\H_7$, the rest of the homology groups are just given by crossing homology classes of the components.  In all we find
\beq
\H_*(\rpt\times\rps)=(\Z,\Z_2^2,\Z_2,\Z\oplus\Z_2^2,\Z_2^2,\Z_2^2,\Z_2^2,
\Z\oplus\Z_2,\Z_2^2,0,\Z) \,.
\eeq
The universal coefficient theorem says that the cohomology groups are isomorphic to the homology groups but with the torsion parts moved up one dimension
\beq
\H^*(\rpt\times\rps)=(\Z,0,\Z_2^2,\Z\oplus\Z_2,\Z_2^2,\Z_2^2,\Z_2^2,\Z\oplus\Z_2^2,
\Z_2,\Z_2^2,\Z)\,.
\eeq

We are not interested in just the additive structure of $\H^*$, but also the multiplicative structure.  As in the case of $\H^*(\rps)$ above, we can learn the multiplicative structure by writing the generators of $\H^*(\rpt\times\rps)$ in terms of the $\Z_2$ cohomology rings of the constituent real projective spaces.  In most cases this will be made easier by the fact that the classes in the product cohomology are just products of the $\Z$ classes of the constituent cohomologies given in Eq.~(\ref{rpgen}).  The exceptions are the three classes $\H^3$, $\H^5$ and $\H^7$ in which the $Tor$ term contributed an extra $\Z_2$ to the homology.  These cohomology classes will correspondingly contain an extra $\Z_2$ factor that is not a product of two cohomology classes of the components, but can be expressed as the Bockstein of a product of two $\Z_2$ cohomology classes $a$ and $b^{2j+1}$ of the constituent $\rpn$'s.   

\begin{table}[h]
\begin{center}
\begin{tabular}{|c|c|c|c|c|c|c|c|c|c|c|}
\hline $\H^0$ & $\H^1$ & $\H^2$ & $\H^3$ & $\H^4$ & $\H^5$ & $\H^6$ & $\H^7$ & $\H^8$ & $\H^9$ & $\H^{10}$\\
\hline $\Z$&$0$&$\Z_2^2$&$\Z\oplus\Z_2$&$\Z_2^2$&$\Z_2^2$&$\Z_2^2$&$\Z\oplus\Z_2^2$&$\Z_2$&$\Z_2^2$&$\Z$\\
\hline $1$&$  $&$a^2 $&$a^3$&$ b^4$&$ a^3b^2 $&$b^6 $&$b^7,a^3b^4$&$a^2b^6   $&$a^3b^6 $&$a^3b^7 $\\
&&$b^2$&$ab^2+a^2b$&$a^2b^2$&$ab^4+a^2b^3$&$a^2b^4$&$ab^6+a^2b^5$&&$a^2b^7$&\\
\hline
\end{tabular}
\end{center}
\caption{\footnotesize  The cohomology groups and their generators are summarized.}
%\label{bbbb}
\end{table}

%As in the $\rpn$ case the product structure of the cohomology ring of $\rpt\times\rps$ is induced from the above decomposition into $\Z_2$ cohomology generators.

% =========================================================================
\subsection{The Untwisted K-Theory of $\rpt\times\rps$}

We will compute the (untwisted) K-theory of $\rpt\times\rps$ using the Atiyah-Hirzebruch spectral sequence (AHSS).  In fact only the first differential of this sequence
\beq
d_3=Sq^3
\eeq
will be nontrivial.  $Sq^3$ is the third Steenrod square.  It can be decomposed as
\beq
Sq^3=\beta \, sq^2 \,,
\eeq
where $\beta$ is the Bockstein map which lifts torsion $p$-cocycles to the $(p+1)$st integral cohomology.  The Bockstein map on $\Z_2$ classes has the following physical interpretation.  Branes in type II string theory are necessarily oriented.  Therefore if a half brane is wrapped on a nonorientable $(p+1)$-chain $Z_{p+1}$ then it must fail to close\footnote{Intuitively the orientation of the brane may be defined everywhere except for a Dirac string, and that Dirac string is the insertion.} on the $p$-cycle $\partial Z_{p+1}$ which, due to the nonorientability of $Z_{p+1}$, is divisible by two.  This means that, in addition to wrapping $Z_{p+1}$, it necessarily extends away from $Z_{p+1}$ making a cylinder whose cross-sections are each $\partial Z_{p+1}$.  The Bockstein is dual to the half of $\partial Z_{p+1}$ that is wrapped by the cylindrical brane, and so the Bockstein map takes a half-brane to the cross-section of the tube of whole-brane that must be inserted to make it consistent (oriented).  The half-brane itself will, in this case, result from the $sq^2$ term.

As a first step we will compute $Sq^3$ on all of the generators of our cohomology ring.  Recall that all of the generators but three are products of integer classes of the cohomology of the individual $\rpn$'s.  These integer classes are either zero classes or top classes, which are annihilated by $sq^2$, or else products of even integral classes.  If they are products of degree two integral classes then we may iteratively use the Cartan rule for Steenrod square two
\beq
sq^2 (ab)=(sq^2 a)b+(Sq^1 a)(Sq^1 b)+a(sq^2b) \,,
\eeq
and the fact that $Sq^1=\beta$ annihilates integral classes, to write $sq^2$ on each class as an integral class times $sq^2$ of an integral 2-class.  $sq^2$ acts on any 2-class by squaring it, and so it yields another integral class.  Thus all of our generators except for possibly the three special ones will be mapped to integral classes ($\Z_2$ classes with integral lifts) by $sq^2$.  

The Bockstein map annihilates $\Z_2$ classes that lift to integral classes of the same dimension.  Thus $Sq^3$ will annihilate all of our generators, except for possibly the three special ones.  One of the special generators, $ab^6+a^2b^5$, is degree 7 and so $Sq^3$ of it will be degree 10.  The image of $Sq^3$ is always mod 2 torsion, but $H^{10}(\rpt\times\rps)$ does not contain any torsion, thus $Sq^3$ will annihilate this generator as well.  We then only need to evaluate $Sq^3$ on the other two special generators.  

We may evaluate $Sq^3$ on the first special generator by using the fact that $Sq^3$ squares three classes.  Thus
\beq
Sq^3(ab^2+a^2b)=(ab^2+a^2b)(ab^2+a^2b)=a^2b^4+a^4b^2=a^2b^4\,. \label{hsq}
\eeq
The second term vanishes because $a^4=0$.  To evaluate $Sq^3$ on the second generator we will also use the Cartan rule for $Sq^3$ and the fact that all of the Steenrod squares annihilate powers of $a$ except for $Sq^1 a=a^2$.
\bea
Sq^3(ab^4+a^2b^3)&=&Sq^3(ab^4)+Sq^3(a^2b^3)=(Sq^1a)sq^2b^4+ aSq^3b^4+a^2Sq^3b^3\nonumber\\
&=&(a^2)[(sq^2b^2)b^2+b^2(sq^2b^2)]+a\beta[(sq^2b^2)b^2+b^2(sq^2b^2)]+a^2b^6\nonumber\\
&=&a^2[2b^6]+a\beta[2b^6]+a^2b^6=0+0+a^2b^6=a^2b^6.
\eea

We may now use the AHSS to compute the associated graded part of the untwisted K-theory of $\rpt\times\rps$.   It is just given by the cohomology of $Sq^3$
\bea
Gr(K^0)&=&\frac{Ker(Sq^3):H^{\even}\longrightarrow H^{\odd}}{Im(Sq^3):H^{\odd}\longrightarrow H^{\even}}=\frac{\Z^2\oplus\Z_2^{7}}{\Z_2^2}=\Z^2\oplus\Z_2^{5}  \,, \nonumber\\
Gr(K^1)&=&\frac{Ker(Sq^3):H^{\odd}\longrightarrow H^{\even}}{Im(Sq^3):H^{\even}\longrightarrow H^{\odd}}=\frac{\Z^2\oplus\Z_2^5}{0}=\Z^2\oplus\Z_2^{5}\,.  \label{associato}
\eea
To calculate the K-theory ring from the associated graded K-group one needs to solve an extension problem, which in this case allows each $\Z$ to eat either zero or some $\Z_2$'s and allows the $\Z_2$'s to combine into $\Z_{2^j}$'s.  We may alternatively find the K-theory by T-dualizing the configuration to $S^2\times \cp^3\times T^2$ with $H$-flux, explicitly constructing the bundles on $\cp^3$, and noting that two vortices make an instanton while two instantons make a codimension 6 instanton, which corresponds to the assembly of the three corresponding $\Z_2$'s into a $\Z_8$.  

Instead, in the spirit of Ref.~\cite{Bra}, we will solve for the untwisted K-theory of $\rpt\times\rps$ by using the K\"unneth formula for K-homology
\beq
0\longrightarrow K_*(A)\otimes K_*(B)\longrightarrow K_*(A\times B)\longrightarrow \textup{Tor}(K_*(A),K_*(B))\longrightarrow 0
\eeq
as well as the untwisted K-homologies of the constituents
\beq
K_0(\rpt)=\Z\hsp K_1(\rpt)=\Z\oplus\Z_2\hsp K_0(\rps)=\Z\hsp K_1(\rps)=\Z\oplus\Z_8\,.
\eeq
Using the fact that $\Z_2\otimes\Z_8=\Z_2$ and that $Tor(x,y)$ vanishes unless both $x$ and $y$ contain torsion components we find that
\bea
K_0(\rpt\times\rps)&=&\big(K_0(\rpt)\otimes K_0(\rps)\big)
\oplus \big( K_1(\rpt)\otimes K_1(\rps) \big)\\
&&\oplus\textup{Tor}(K_0(\rpt),K_1(\rps))\oplus\textup{Tor}(K_1(\rpt),K_0(\rps))\nonumber\\
&=&\big(\Z\otimes\Z\big)
\oplus \big( (\Z\oplus\Z_2)\otimes(\Z\oplus\Z_8)\big) \oplus 0 \oplus 0=\Z^2\oplus\Z_8\oplus\Z_2^2\,.\nonumber\\
\eea
while $Tor(\Z_2,\Z_8)=\Z_2$ yields
\bea
K_1(\rpt\times\rps)&=&\big( K_0(\rpt)\otimes K_1(\rps)\big) \oplus \big(K_1(\rpt)\otimes K_0(\rps) \big) \nonumber\\
&&\oplus\textup{Tor}(K_0(\rpt),K_0(\rps))\oplus\textup{Tor}(K_1(\rpt),K_1(\rps))\nonumber\\
&=&\big( \Z\otimes(\Z\oplus\Z_8)\big) \oplus \big( (\Z\oplus\Z_2)\otimes\Z\big)
\oplus 0 \oplus
Tor(\Z_2,\Z_8)\nonumber\\
&=&\Z^2\oplus\Z_8\oplus\Z_2^2 \,.
\eea
The universal coefficient theorem then yields the desired K-cohomology groups
\beq
K^0(\rpt\times\rps)=\Z^2\oplus\Z_8\oplus\Z_2^2\hsp
K^1(\rpt\times\rps)=\Z^2\oplus\Z_8\oplus\Z_2^2\,. \label{kappasenzaacha}
\eeq
These K-theory classes yield flux backgrounds or brane configurations for type IIA or IIB string theory.  In IIB string theory the $K^1$ classes describe possible RR fields in the absence of an NS flux, and if we S-dualize a configuration in which the $G_3$-flux valued in $\Z\subset \H^3(\rpt\times\rps)$ is nonzero then we find another configuration which corresponds to a class in the K-theory twisted by the original $G_3$.

The K-homology groups are also both $\Z^2\oplus\Z_8\oplus\Z_2^2$.  In type IIB the $\Z$'s are generated by the $D(-1)$ and the $D9$.  The $\Z_8$ is generated by a D7 wrapping $\rpt\times\rp^5$.  Similarly to the case of Sec.~\ref{riscaldimento}, the element $2\in\Z_8$ corresponds to a D5 wrapping $\rpt\times\rpt$ and 4 to a D3 wrapping $\rpt\times\rp^1$.  The $\Z_2$ generators are the D7 wrapping $\rp^1\times\rps$ and the D5 wrapping $\rp^1\times\rp^5$.  One might be tempted to think that the D5 wrapping $\rp^1\times\rp^5$ would be $\Z_8$ charged like its cousin that wraps $\rp^3\times\rp^5$.  Above we have seen that the $\Z_2$ charge results from the fact that $\Z_2\otimes\Z_8=\Z_2$.  Physically this is a consequence of the instability of the element $2$, which is the D3 wrapping $\rp^1\times\rp^3$.  The D3 may split into two D5's which each wrap $\rp^1\times\rp^5$ and carry a half unit of D3 charge.  These two D5's each wrap the circle $\rp^1\subset\rp^3$ and they may connect so as to wrap the circle $\rp^1\subset\rp^3$ twice.  However twice this circle is the boundary of a $\rp^2\subset\rp^3$, that is $\partial\rp^2=2\rp^1$.  Thus the doubly-wrapped D5 may decay by sweeping out $\rp^2$.  This entire process is mathematically just the distributive property of the tensor product
\beq
(1\in\Z_2)\otimes(2\in\Z_8)=(2\in\Z_2)\otimes(1\in\Z_8)=0\otimes(1\in\Z_8)=0.
\eeq
We summarize the wrappings and the corresponding brane charges as elements of $\Z\oplus\Z\oplus\Z_8\oplus\Z_2\oplus\Z_2$ in the following table, where the top row is the element of $K_0$ and the bottom row is the subset of $\rp^3\times\rp^7$ wrapped.
\begin{table}[h]
\begin{center}
\begin{tabular}{|c|c|c|c|c|c|c|}
\hline $(1,0^4)$ & $(0,1,0^3)$ & $(0^2,1,0^2)$ & $(0^2,2,0^2)$ & $(0^2,4,0^2)$ & $(0^3,1,0)$ & $(0^4,1)$\\
\hline a point&$\rpt\ptimes\rps$&$\rpt\ptimes\rp^5$&$\rpt\ptimes\rpt$&$\rpt\ptimes\rp^1$&$\rp^1\ptimes\rp^7$&$\rp^1\ptimes\rp^5$\\
\hline
\end{tabular}
\end{center}
\caption{\footnotesize  Some K-homology classes and the corresponding wrapped cycles}
%\label{bbbb}
\end{table}

Again one may replace $\rps$ by $\rp^{2k+1}$ by replacing the $\Z_8$'s by $Z_{2^k}$'s, as there will be only $k$ $\Z_2$'s.  The terms with $b^{j>2k+1}$ are no longer present as they do not fit in $\rp^{2k+1}$.  Crossing the spacetime by an even-dimensional sphere just corresponds to doubling the K-groups $K\mapsto K\oplus K$.

% =========================================================================
\subsection{The Twisted K-Theory}

We are interested in configurations that do not correspond to S-covariant classes.  To construct such a configuration we turn on an $H$-flux valued in the $\Z_2$ part of $\H^3(\rpt\times\rp^{2k+1}\times S^{6-2k})$, that is
\beq \label{eqHflux}
H=ab^2+ba^2\,.
\eeq
If $k>1$ (so that $b^4\neq 0$) this will not yield an S-covariant configuration because, as we have just seen in Eq.~(\ref{hsq}),
\beq
Sq^3 H=a^2b^4\neq 0\,.
\eeq
However a twisted K-theory exists with every possible twist in $\H^3$ with $\Z$ coefficients and so it does make sense to calculate the twisted K-theory $K_H(\rpt\times\rps)$, so long as we remember that fluxes valued in this group will lead to anomalies that need to be canceled.  If $k=3$ then
\beq
H\cup H\cup H = a^3b^6\neq 0 \,,
\eeq
which, as was argued in the introduction, implies that unless $H\cup P=H\cup H\cup H$ the anomaly cannot be canceled and so the $k=3$ case will be unphysical.  We have not been able to calculate $P$, however if indeed $P$ does contain a $a^2b^4$ term and so render the $k=3$ case consistent then the D3-brane dual to $a^2b^4$ will not appear in the twisted case but instead will appear in the untwisted case.  T-dualizing the untwisted case, as we will see in the next section, we arrive at a spacetime with no torsion and so the T-dual of this gravitational D3-brane should be a D-string produced by a Freed-Witten anomaly of the fluxes.  One might hope to compute $P$ by examining the Freed-Witten anomalies of these dual fluxes, but the result is heavily dependent on factors of two that we have so far been unable to determine.  A D-string analogue of our proposed formula for D3-brane charge would be helpful.

The associated graded twisted K-theory is in this case just the cohomology with respect to the differential
\beq
d_3=Sq^3+H \,,
\eeq
and so we need to calculate the cup product of $H$ with all of our cohomology generators.  Like $Sq^3$, $H\cup$ will lead to a torsion class three degrees higher and so $H$ will automatically annihilate all of the classes of degree seven and above.  Similarly $H$ will annihilate any term containing $a^3$, as $a^3$ is already of the maximum degree in $\H^*(\rpt)$. The products with $H$ are computed using multiplication in the respective $\Z_2$ rings and then lifting the results to integral cohomology.  We find
\bea
&&H\cup1=ab^2+a^2b\hsp H\cup a^2=a^3b^2\hsp H\cup b^2=ab^4+a^2b^3\,,\nonumber\\
 &&H\cup(ab^2+a^2b)=a^2b^4\hsp H\cup a^2b^2=a^3b^4\hsp H\cup b^4=ab^6+a^2b^5\,,\nonumber\\
&& H\cup(ab^4+a^2b^3)=a^2b^6\hsp H\cup a^2b^4=a^3b^6\hsp H\cup b^6=a^2b^7\,.
\eea
Note that while $Sq^3$ and $H$ are both nontrivial on the special 3 and 5 cocycles $H$ and $H\cup b^2$, $d_3$ annihilates them both.  However they are both in the image of $d_3$ and so they will be quotiented out of the final answer.

We may now assemble the above results to find the associated graded twisted K-theory.
\bea
Gr(K^0_H)&=&\frac{Ker(d_3):H^{\even}\longrightarrow H^{\odd}}{Im(d_3):H^{\odd}\longrightarrow H^{\even}}=\frac{\gnr{2,a^2b^6,a^3b^7}}{0}=\Z^2\oplus\Z_2\nonumber\\
Gr(K^1_H)&=&\frac{Ker(d_3):H^{\odd}\longrightarrow H^{\even}}{Im(d_3):H^{\even}\longrightarrow H^{\odd}}\nonumber\\
&=&\frac{\langle a^3,b^7,ab^2+a^2b,ab^4+a^2b^3,a^3b^2,a^3b^4,ab^6+a^2b^5,a^3b^6,a^2b^7 \rangle}{\langle{ab^2+a^2b,ab^4+a^2b^3,a^3b^2,a^3b^4,ab^6+a^2b^5,a^3b^6,a^2b^7}\rangle}\nonumber\\
&=&\frac{\Z^2\oplus\Z_2^7}{\Z_2^7}=\Z^2.
\eea

$Gr(K^1_H)$ has no torsion and so the extension problem for $K^1_H$ is trivial and we may conclude that
\beq
K^1_H(\rpt\times\rps)=\Z^2\,.
\eeq
However to find $K^0_H$ we need to solve the extension problem
\beq
\Z^2\stackrel{f}{\longrightarrow} K^0_H(\rpt\times\rps)\stackrel{g}{\longrightarrow}\Z_2\,. \label{eprob}
\eeq
This admits two possible solutions.  If the first map, $f$, were multiplication by one then the sequence would split and $K^0_H$ would be $\Z^2\oplus\Z_2$.  If on the other hand $f$ were multiplication by two on one $\Z$ then $K^0_H$ would be $\Z^2$.  A very similar extension problem occurs in the computation of the twisted K-theory of $\rp^{2k+1}\times S^1$.  There it has been shown by T-duality \cite{BEMa} that the map is indeed multiplication by two.  We will argue in the next section that the configuration $\rp^3\times \rp^7$ with H-flux \eqref{eqHflux}
is T-dual to $S^3\times S^7$ with no H-flux, for which the two K-theory groups are both $\Z^2$.  Thus the fact that T-duality is an isomorphism of twisted K-theory \cite{RR,BEMa} will allow us to to conclude that this time, as in the case of $\rp^{2k+1}\times S^1$, $f$ is multiplication by two and so
\beq
K^0_H(\rpt\times\rps)=\Z^2\,.
\eeq
In Sec.~\ref{inserzioniduale}, when we explicitly T-dualize the brane dual to the $\Z_2$ cycle $a^2b^6$, we will give a physical interpretation of the fact that $f$ is degree two.  

The torsion has all been killed and so the twisted K-groups are $k$-independent.  However note that all K-groups double to $\Z^4$ when we cross our $\rpt\times\rp^{2k+1}$ with a sphere.

% =========================================================================
\section{T-duality} \label{t-dualita}

\subsection{The Untwisted Case: $\cp^1\times\cp^3\times T^2$}

As a warm up for the more difficult T-duality to come, we dualize the $k=3$ example of $\rpt\times\rps$ with no $H$-flux.  Both $\rpt$ and $\rps$ admit free circle actions and in particular are circle bundles over the complex projective spaces $\cp^1$ and $\cp^3$ respectively
\begin{equation}
\begin{CD}
{S^1} @>>> \rp^{2j+1}\\
&& @V\hat{\pi} VV \\
&& \cp^j  \label{rp7} \end{CD} 
\end{equation}
The nontrivial cohomology classes and generators of the base spaces are
\bea
\H^0(\cp^1)=\H^0(\cp^3)=\Z=\gnr{1}\hsp
\H^2(\cp^1)=\Z=\gnr{\a}\,, \nonumber\\
\H^2(\cp^3)=\Z=\gnr{\b}\hsp
\H^4(\cp^3)=\Z=\gnr{\b^2}\hsp
\H^6(\cp^3)=\Z=\gnr{\b^3}\,,
\eea
and the Chern classes of the two bundles are
\beq
c_1=2\a\,\hspace{.5cm}\textup{and}\hspace{.5cm} c_1=2\b\,.
\eeq

We may now use the prescription of Ref.~\cite{BEMa} to T-dualize both circle fibers. According to that prescription T-duality exchanges the integrals of the $H$-fluxes over the circle fibers with the Chern classes.  The original $H$-flux vanishes and so the dual Chern classes vanish
\beq
\hat{c}_1=0\,.
\eeq
The dual spacetime $\widehat{M}$ then consists of the product of two trivial circle bundles over the original base
\beq
\widehat{M}=\cp^1\times\cp^3\times S^1_\a\times S^1_\b.  
\eeq
The dual $H$-flux is just the sum of two $H$-fluxes that integrate to the Chern classes
\beq
\widehat{H}=2\a\cup \tha+2\b\cup \thb\,,
\eeq
where $\tha$ and $\thb$ generate $H^1(S^1,\Z)$ of the dual circles.

We will now compute the twisted K-theory of $\widehat{M}$ which, as we have performed an even number of T-dualities, must agree with the untwisted K-theory of the original space given in Eq.~(\ref{kappasenzaacha}).  The cohomology of $\widehat{M}$ follows from the K\"unneth theorem, where the $Tor$ terms all vanish as there is no torsion.
\begin{table}[h]
\begin{center}
\begin{tabular}{|c|c|c|c|c|c|c|c|c|c|c|}
\hline $\H^0$ & $\H^1$ & $\H^2$ & $\H^3$ & $\H^4$ & $\H^5$ & $\H^6$ & $\H^7$ & $\H^8$ & $\H^9$ & $\H^{10}$\\
\hline $\Z$&$\Z^2$&$\Z^3$&$\Z^4$&$\Z^4$&$\Z^4$&$\Z^4$&$\Z^4$&$\Z^3$&$\Z^2$&$\Z$\\
\hline $1$&$\tha$&$\a$&$\a\tha$&$\a\tha\thb$&$\a\b\tha$&$\a\b\tha\thb$&$\a\b^2\tha$&$\a\b^2\tha\thb$&$\a\b^3\tha$&$\a\b^3\tha\thb$\\
\hline $$&$\thb$&$\b$&$\a\thb$&$\a\b$&$\a\b\thb$&$\a\b^2$&$\a\b^2\thb$&$\a\b^3$&$\a\b^3\thb$&$$\\
\hline&$$&$\tha\thb$&$\b\tha$&$\b\tha\thb$&$\b^2\tha$&$\b^2\tha\thb$&$\b^3\tha$&$\b^3\tha\thb$&$$&\\
\hline&$$&$$&$\b\thb$&$\b^2$&$\b^2\thb$&$\b^3$&$\b^3\thb$&$$&$$&\\
\hline
\end{tabular}
\end{center}
\caption{\footnotesize  The cohomology groups of $\widehat{M}$ and their generators are summarized.}
\label{mcappuccio}
\end{table}
The fact that the cohomology contains no torsion also implies that $Sq^3$ annihilates everything.  The cup product with $H$ acts as follows
\bea
&&1\hcup 2\a\tha\pic+\pic2\b\thb\hsp
\tha\hcup -2\b\tha\thb\hsp
\a\hcup 2\a\b\thb\hsp
\a\tha\hcup -2\a\b\tha\thb,\nonumber\\
&&\b\hcup 2\a\b\tha\pic+\pic2\b^2\thb\hsp
\b\tha\hcup -2\b^2\tha\thb\hsp
%b\thb\hcup 2ab\tha\thb\hsp
\a\b\hcup 2\a\b^2\thb\hsp
\a\b\tha\hcup -2\a\b^2\tha\thb,\nonumber\\
&&\b^2\hcup 2\a\b^2\tha\pic+\pic2\b^3\thb\hsp
\b^2\tha\hcup -2\b^3\tha\thb\hsp
\a\b^2\hcup 2\a\b^3\thb\hsp
\a\b^2\tha\hcup -2\a\b^3\tha\thb,\nonumber\\
&&\b^3\hcup 2\a\b^3\tha,
\thb\hcup 2\a\tha\thb,
\b\thb\hcup 2\a\b\tha\thb,
\b^2\thb\hcup 2\a\b^2\tha\thb,
\b^3\thb\hcup 2\a\b^3\tha\thb,
\nonumber\\
&&\a\thb,
\a\b\thb,
\a\b^2\thb,
\a\b^3\thb,
\a\tha\pic+\pic\b\thb,
\a\b\tha\pic+\pic\b^2\thb,  
\a\b^2\tha\pic+\pic\b^3\thb,
\a\b^3\tha,
\b^3\tha\hcup 0\,,\nonumber\\
&&\a\b^3,
\tha\thb,
\b\tha\thb,
\b^2\tha\thb,
\b^3\tha\thb,
\a\tha\thb,
\a\b\tha\thb,
\a\b^2\tha\thb,
\a\b^3\tha\thb\hcup 0\,.
\eea

The first approximation to $K^i$, which we will call $E_1^i$, is the even or odd part of the quotient of the kernel of $d_3=H\cup$ by the image of $d_3=H\cup$.  Inspecting the above action of $H$ we see that
\beq
E_1^0=E_1^1=\Z^2\times\Z_2^7\,.
\eeq
These have bigger torsion subgroups, each with 128 elements, than the T-dual K-classes, which each had only 32.  While the extension problem could in principle resolve this discrepancy, we will argue that instead the extra classes are removed by the higher AHSS differential $d_5$.  These will remove precisely the same elements that were removed, in the derivation of the T-dual K-theory using the K\"unneth theorem, by the fact that
\beq
\Z_8\otimes \Z_2=Tor(\Z_8,\Z_2)=\Z_2\,,
\eeq
which is smaller than the original $\Z_8$ by a factor of 4 \cite{Bra}.

To find the action of $d_5$ we will use the dual description of these cohomology classes in terms of D-branes.  The product of two T-dualities is an isomorphism of twisted K-theory and furthermore the extension problem is the same as before, yielding the same pattern of fractional brane charges inside of higher dimensional branes.  In particular, the branes wrapping the $\cp^2\subset\cp^3$ each carry half a unit of charge of a brane wrapping $\cp^1\subset\cp^3$ while those wrapping $\cp^1\subset\cp^3$ each carry half a unit of charge of a brane at a point in the $\cp^3$ directions.  The first of these facts follows from the fact that the normal bundle of $\cp^2\subset\cp^3$ is not $spin$, and the second follows from this argument with the same factor of 2 that we saw in the K-theory of $\rps$.  In terms of the Steenrod algebra these two relations are consequences of
\beq
sq^2\, \b=\b^2\hsp P^1 \b = \b^3,
\eeq
where again $P^1$ acting on 2-torsion is a secondary operation that
cubes 2-classes.  In fact the $\rp^7$ case is just a pullback of this one using the projection map $\pi:\rp^7\longrightarrow \cp^3$ of Eq.~(\ref{rp7}). 

\begin{figure}[ht]
\begin{center}
\leavevmode
\epsfxsize 10   cm
\epsffile{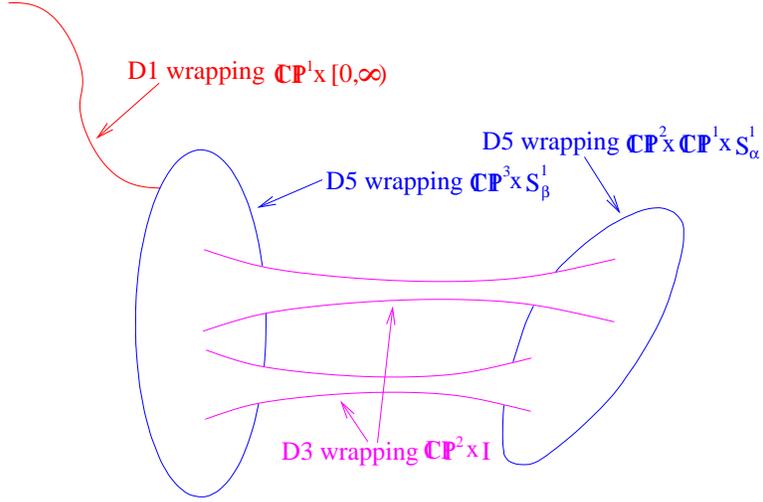}    
\end{center} 
\footnotesize{\caption{Two D5 components have worldvolume $H$-fluxes that cancel, meaning that one absorbs the D3's sourced by the other.  Thus the combination is $d_3$ closed, but it is not $d_5$ closed.  This means that the pair of D5's requires a D1 insertion, whose cross-sections are dual to $d_5$ of the cocycle dual to the D5 pair.  D$(p+4)$-D$p$ baryons are always dual to cycles related by $d_5$.}}
%\label{2dmu} 
\end{figure}

Consider for example the D5-brane dual to $\a\tha+\b\thb$.  This brane consists of two components, one wrapping $\cp^3\times S^1_\b$ and the other wrapping $\cp^1\times\cp^2\times S^1_\a$, where the extra dimension again comes from the fact that dual branes live in the space augmented by an extra deformation direction.   Both of these components contain $H$-flux, but the orientations disagree and the total integral of $H$ cancels
\beq
\int_{\cp^3\times S^1_\b} H = -\int_{\cp^1\times\cp^2\times S^1_\a} H=2\int_{\cp^2}\,. \label{cancelazione}
\eeq
Here the integral denotes the homology-cohomology pairing over the integers, and the terms are to be interpreted as operators that act on cohomology over the integers. 

The $H$-flux on each component is nonvanishing, and so each component requires a D3-brane insertion to cancel the corresponding Freed-Witten anomaly, or equivalently, as a sink for the worldvolume magnetic flux sourced by the $H$-flux.  The cancellation (\ref{cancelazione}) means that the same number of D3-branes are emitted from one D5 as are absorbed by the other and so this is not a D5-D3 baryon, reflecting the fact that $a\tha+b\thb$ is $d_3$-closed.  As there are two units of $H$-flux, two D3-branes connect the pair of D5's.  To sink the magnetic flux sourced by $H$, the intersection of the pair of D3's and each D5 component needs to be Poincar\'e dual to the $H$-flux restricted to the components worldvolume.  This means that each D3 wraps $\cp^2\subset\cp^3\times S^1_\b$ in the first component and $\cp^2\subset\cp^1\times\cp^2\times S^1_\a$ in the second.  It is a critical test of the consistency of this picture that the topology of the tube of D3-brane is the same at both endpoints.

As the normal bundle of $\cp^2\subset\cp^3$ is not $spin$, the D5
component wrapping $\cp^1\times\cp^2\times S^1_\a$ must contain a half
unit of charge of a D3-brane wrapped on $\cp^1\times\cp^1\times
S^1_\a$, which is dual to $w_2$ of its normal bundle.  Similarly the
D3-brane tube wraps $\cp^2\subset\cp^3$ times an interval and so it
must contain a half unit of D1 charge on $\cp^1\subset\cp^3$ times the
interval.  On the other hand the second component $\cp^3\times
S^1_\b$, which has a trivial normal bundle, apparently supports no D3
charge.  Thus the D1-charge inside of the D3-brane cannot end on the
second component, and must instead continue off to infinity.  There
are two D3 tubes each carrying half a unit of D1 charge, and so there
is a total of 1 unit of D1 charge escaping.  Our entire configuration
is then a D1-D5 baryon.  While each D3 carried half-integral D1
charge, the D1-charge escaping needed to be integral to satisfy the
Dirac quantization condition.  The fact that it indeed is integral
corresponds to the existence of the division by two in the definition
of $P^1$.

As in Refs.~\cite{MMS, Uday} a D$p$-D$(p-4)$ baryon implies that $d_5$ of the dual of the D$p$ brane is the dual of its intersection with the D$(p-4)$ brane.  In our case this reads
\beq
d_5 (\a\tha+\b\thb)= \a\b^2\tha\thb\,.
\eeq
Notice that this agrees with the formula for $d_5$ conjectured in Ref.~\cite{Uday}
\bea
d_5(\a\tha+\b\thb)&=&\left( 
sq^2\,\frac{H}{2} 
\right) (\a\tha+\b\thb)=(sq^2(\a\tha+\b\thb))(\a\tha+\b\thb)\nonumber\\
&=&\b^2\thb(\a\tha+\b\thb)=\a\b^2\tha\thb\,.
\eea
As always we may interpret the two factors of the differential as the two steps connecting the D1 and the D5.  First the $H$ takes the D5-brane to the two D3 tubes, then the $sq^2/2$ calculates the D1 charge of each tube.  This is the same factor of $2$ that came into the two step brane within brane embedding in the solution of the extension problem for $\rp^7$.  Perhaps when there are $n$ steps of embedding there is always a factor of $(n!)$ that counts the orderings of the embeddings.

This story proceeds analogously if we multiply our class by a factor of $\b$, so that each brane that wrapped $\cp^j\subset\cp^3$ now wraps $\cp^{j-1}\subset\cp^3$.  We then find
\beq
d_5(\a\b\tha+\b^2\thb)=\left( sq^2\,\frac{H}{2} \right)
(\a\b\tha+\b^2\thb)=\b^2\thb(\a\b\tha+\b^2\thb)=\a\b^3\tha\thb \,,
\eeq
which again is in agreement with Ref.~\cite{Uday}.

Restricting to the kernel of $d_5$ now kills two of the seven factors of $\Z_2$,
and another two factors are killed when we quotient by the image.  In
all we lose two $\Z_2$'s from each associated graded K-group and thus the quotient of the kernel of $d_5$ by its image is the same associated graded K-group found for its T-dual in Eq.~(\ref{associato}).  The extension problems proceed identically, and so the twisted K-theory of $\widehat{M}$ agrees with the untwisted K-theory of its T-dual $\rpt\times\rps$.  Note that here the secondary operation, $d_5$, is T-dual to the primary operation $d_3$.  Secondary operations are notoriously difficult to calculate, but this example suggests the possibility that in some classes they may be calculable as primary operations on an auxiliary space.  For example one may try to find a recursive relation satisfied by Massey products.

% =========================================================================
\subsection{The Twisted Case: $S^3\times S^{2k+1}\times S^{6-2k}$}

$\rp^{2k+1}$ is a circle bundle over $\cp^k$ with Chern class equal to two.  Thus $\rpt\times\rp^{2k+1}\times S^{6-2k}$ is a 2-torus bundle over $\cp^1\times\cp^k\times S^{6-2k}$.  We claim that  T-dualizing two particular generators of this torus, in the presence of the above torsion $H$-flux
(Eq.\ \eqref{eqHflux}), yields IIB string theory on $S^3\times S^{2k+1}\times S^{6-2k}$ with no $H$-flux.  We will ignore the spheres $S^{6-2k}$ which support neither the $H$-flux nor the curvature of the circle bundles. 

We begin by T-dualizing the circle $S^1_a$ fiber in $\rpt$
\begin{equation}
\begin{CD}
S^1_a @>>> \rpt\times\rp^{2k+1} \\
&& @V\pi VV \\
&& \cp^1\times\rp^{2k+1} \end{CD}
\end{equation}
which has Chern class $2\alpha$, where $\alpha$ is the generator of  $\H^2(\cp^1)=\Z$.  The circle $S^1_a$ is trivially fibered over the $\rp^{2k+1}$, but the $\rp^{2k+1}$ is included in this diagram because the T-dual circle $\hsa$ will be nontrivially fibered over it.  As always the T-dual fibration is defined by setting the first Chern class equal to the pushforward of the original $H$-flux
\begin{equation}
{\widehat F}=\pi_! H = \pi_! (ab^2+a^2b)=\int_{S^1_a} H\,.
\end{equation}
The pushforward is just the integral of a differential form representative of $H$ (which by an abuse of notation we have also called $H$ above) over the fiber $S^1_a$.  Although this three-form is exact, it is the derivative of a two-class that does not satisfy the quantization condition (it is half-integral) and so this integral will give a two-form on $S^2\times\rp^{2k+1}$ which is again the derivative of a form that does not satisfy the quantization condition.  There are many differential form representatives of $H$ that differ by exact forms that are derivatives of forms that do satisfy the quantization condition, for example three times the representative chosen.  However we will see that the answer is a $\Z_2$ cohomology class and so is unaffected by the addition of even classes.

While we could construct an explicit representative and do the integral, an easier approach is to evaluate the pushforward by equating it with the homology-cohomology pairing with $\Z_2$ coefficients.\footnote{Alternatively the pushforward of $H$ may be calculated by using the exactness of the Gysin sequence for this circle bundle.}
Then the pushforward is just the pairing with the generator $\langle a |$ of $H_1(\rpt;\Z_2)$, which is our fiber.  We then lift our result back to the integral cohomology of the base $\cp^1\times\rp^{2k+1}\times S^{6-2k}$.  Thus
\begin{equation}
{\widehat F}=\pi_!(ab^2+a^2b)=\langle a|ab^2+a^2b\rangle=\langle a|ab^2\rangle+\langle a|a^2b\rangle=b^2+2ab=b^2 \,,
\end{equation}
where $2ab$ vanishes because $a$ is $\Z_2$-valued.  The vanishing of this term was necessary because $ab$ does not lift to an integral class in $\H^2(\rpt\times\rp^{2k+1})$.  $b^2$ is the nontrivial class in $\H^2(\rp^{2k+1})=\Z_2$ and so our dual circle $\hsa$ is nontrivially fibered over $\rp^{2k+1}$ and is trivially fibered over the $\cp^1\times S^{6-2k}$
\begin{equation}
\begin{CD}
{\hsa} @>>> \cp^1\times S^{2k+1}\times \hsd\times S^{6-2k}\\
&& @V\hat{\pi} VV \\
&& \cp^1\times \rp^{2k+1}\times S^{6-2k}\ . \end{CD}
\end{equation}
Here we have used the fact that the unique nontrivial $\hsa$ bundle over $\rp^{2k+1}$ is topologically just $S^{2k+1}\times\hsd$.

\begin{figure}[ht]
\begin{center}
\leavevmode
\epsfxsize 10   cm
\epsffile{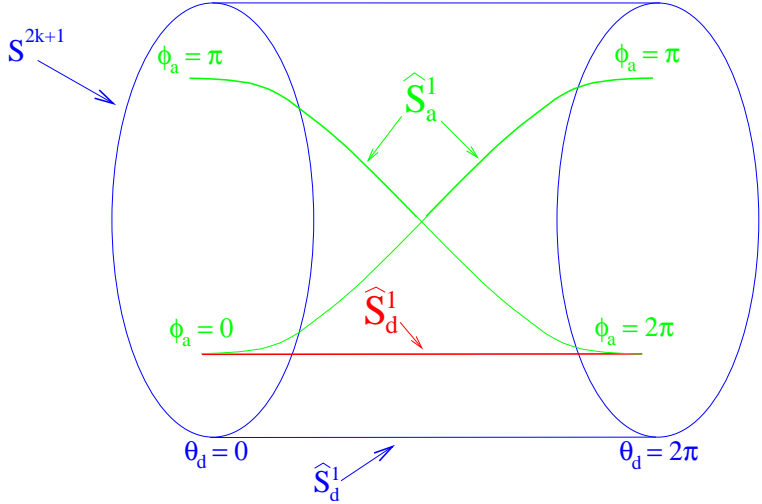}    
\end{center} 
\footnotesize{\caption{The space $S^{2k+1}\times\hsd$ admits two different free circle actions.  One circle action simply goes around the $\hsd$ and so has a space of orbits $S^{2k+1}$.  The other has the same action plus the free circle action on the $S^{2k+1}$, so by the time it has gone around $\hsd$ once it is on the other side of $S^{2k+1}$.  This orbit, called $\hsa$, therefore does not close until it has gone around $\hsd$ twice, when it has returned to its starting point on the sphere.  At a given point on $\hsd$ the orbit $\hsa$ intersects two antipodal points on the $S^{2k+1}$, and so the space of $\hsa$ orbits is the space of pairs of antipodal points, $\rp^{2k+1}$.}}
%\label{2dmu} 
\end{figure}

The relationship between the two circles $\hsa$ and $\hsd$ will be
crucial later and so we will describe it.  The $\hsd$ action may be
seen in the $\hsa$ bundle picture as the diagonal action of $\hsa$
and the free circle action whose orbits are each $\rp^1\in\rp^{2k+1}$.
This diagonal action is free and has a space of orbits $S^{2k+1}$
instead of the original $\rp^{2k+1}$.   Each orbit of $\hsa$ is twice
as long as an orbit of $\hsd$, because after circumnavigating one
$\hsd$ orbit (which would take it back to where it started) the
$\hsa\propto\hsd+\rp^1$ orbit has also traveled around the lift of
the noncontractible loop $\rp^1$, and so it is on the opposite side of sphere\footnote{This is the reason that antipodal points on the sphere are identified in the space of $\hsa$ orbits.}.  Thus the $\hsa$ orbit only closes after the second loop.  An $\hsa$ orbit then consists of two $\hsd$ orbits plus two trips around $\rp^1$.  The $\rp^1$ generates $\H_1(\rp^{2k+1})=\Z_2$ and so two trips around the $\rp^1$ may be deformed into none.  We could have predicted in advance that, except for the factor of 2, our two loops would be homotopic because the fundamental group of $S^{2k+1}\times S^1$ is one-dimensional.

The dual $H$-flux, $\widehat{H}$, is determined from the curvature of the original $S^1_a$ bundle, which had a Chern class of 2 times the generator $\a$ of $\H^2(\cp^1)=\Z$, via the condition \cite{BEMa}
\beq
2\a=F=\hat{\pi}_! \widehat{H}=\int_{\hsa}\widehat{H} \,,
\eeq
and the fact that the parts of $H$ and $\widehat{H}$ in the cohomology of the base $\cp^1\times\rp^{2k+1}$ must agree.  The original $H$ did not contain any terms in the cohomology of the base, and so neither will the dual $\widehat{H}$.  Thus $\widehat{H}$ is simply
\beq
\widehat{H}=2\a\cup \theta_a \,,
\eeq
where $\theta_a$ generates $H^1(\hsa)=\Z$.

This is the answer, but to use the formalism of Ref.~\cite{BEMa} to
perform the next T-duality we will rewrite $\widehat{H}$ in terms of $\hsd$, whose first cohomology generator will be named $\theta_d$.  As \hsa\ is homotopic to twice \hsd\ the cohomology generator $\theta_d$ must be cohomologous to twice the generator $\theta_a$.  This factor of two is necessary to preserve the homology-cohomology pairing, or equivalently the fact that the cohomology classes need to integrate to one over their corresponding cycles.

Note that while as a cohomology class
\beq
[\theta_d]\sim 2[\theta_a] \,,
\eeq
it is in fact crucial, even to the topology of the space of orbits of the circle action, that they differ by the single unit free circle action on the $S^{2k+1}$, or equivalently by two units of the generator of $\H_1(\rps)$.

Summarizing, the first T-duality has left us with $\cp^1\times S^{2k+1}\times\hsd\times S^{6-2k}$ with an NS-flux
\beq
\widehat{H}=2\a\cup \theta_a=\a\cup \theta_d+\a d\phi \sim \a\cup \theta_d \,,
\eeq
where $d\phi$ is the cohomologically trivial unit generator of the
free circle action on the $S^{2k+1}$.  The last equality is an
equality of the cohomology classes that follows from the cohomological
triviality of $d\phi$, but we will return to the $d\phi$ term later.
Note that while shifting $H$ by an exact form $dB$ does not change its
topology, it does change the gauge-invariant Wilson loops $\int B$ and
so may have an effect on the physics.  We have in effect changed the background that we are considering.  We will claim that in the case
$k=3$ a consistent and a potentially inconsistent compactification are related by
such an exact shift.

The second T-duality is much easier.  We now want to T-dualize the circle \hsd, and so we will have T-dualized all of our original $T^2$ bundle over $\cp^1\times\cp^k$, but in a nonorthogonal basis.  If we T-dualized about \hsa\ we would return to where we started.  Despite the fact that \hsa\ is homotopic to a multiple of \hsd\ on the total space, it yields a topologically inequivalent circle bundle\footnote{For example the spaces of orbits, $S^{2k+1}$ and $\rp^{2k+1}$, are not homeomorphic.} and we will see that it leads to a topologically inequivalent T-dual.  However, as the T-duals with respect to \hsa\ and \hsd\ are related to each other by two T-dualities they will necessarily have isomorphic K-theories twisted with respect to their corresponding $H$-fluxes.  %This is because the particular 
%Maldacena-Moore-Seiberg (MMS) instanton that implements 
%the homotopy between the two circles preserves the twisted K-theory class.\footnote{In general MMS instantons each preserve some twisted K-theory inside the S-duality covariant twisted K-theory, but the one preserved may be an S-dual of the one twisted by $H$.}

The circle $\hsd$ is trivially fibered over the base $\cp^1\times
S^{2k+1}\times S^{6-2k}$ with an $H$-flux that contains no component
entirely on the base, thus the dual $H$-flux is trivial.  The
fibration of the dual $\widehat{\widehat{S}^1}$ over $\cp^1\times
S^{2k+1}\times S^{6-2k}$ is described by the Chern class
\beq
\widehat{\widehat{F}}=\int_{\hsd}\widehat{H}=\int_{\hsd} \a \cup \theta_d=\a \,,
\eeq
which is the generator of $\H^2(\cp^1)=\Z$.  This identifies the dual bundle as the Hopf fibration over our 2-sphere, whose total space is the 3-sphere.  The curvature $\widehat{\widehat{F}}$ contains no component on the $(2k+1)$-sphere, and so our resulting spacetime is topologically a Cartesian product of spheres
\begin{equation}
\begin{CD}
{\widehat{\widehat{S}^1}} @>>> S^3\times S^{2k+1}\times S^{6-2k}\\
&& @V\hat{\hat{\pi}} VV \\
&& \cp^1\times S^{2k+1}\times S^{6-2k}\  \end{CD}
\end{equation}
with no $H$-flux.  As there is no $H$-flux the resulting twisted K-theory is just  the untwisted K-theory of $S^3\times S^{2k+1}\times S^{6-2k}$ which is isomorphic to its integral cohomology, as there is no torsion
\beq
K^0(S^3\times S^{2k+1}\times S^{6-2k})=K^1(S^3\times S^{2k+1}\times S^{6-2k})=\Z^2\oplus\Z^2=\Z^4.
\eeq

Note the crucial role played by the factor of two difference in the two circles \hsa\ and \hsd.  Had the corresponding 1-cycles not had a relative factor of two difference in their normalization then there would have been two units of $\widehat{H}$ and so two units of $\widehat{\widehat{F}}$ yielding $\rpt\times S^{2k+1}\times S^{6-2k}$ which would have
\beq
K^0(\rpt\times S^{2k+1}\times S^{6-2k})=\Z^4\oplus\Z_2^2\,.
\eeq
We would not have been able to eliminate this possibility using the AHSS computation of the original twisted K-theory as it would correspond to the other solution of the extension problem (\ref{eprob}).  Thus the first map $f$ of the extension seems to be multiplication by two because, like in the case of the computation of the twisted K-theory of bundles over a single \rpn, the T-dual orbit (\hsa) is twice as long as the shortest orbit (\hsd).  This new shortest element is not in the image of $f$, which only arrives at multiples of \hsa.  Thus we may think of the short element $\hsd$ as the nontrivial element in the image of $g$, $1\in\Z_2$.  The extension would have split, yielding the other result for the twisted K-theory of $\rpt\times\rp^{2k+1}\times S^{6-2k}$ if instead there had been no relation between the loops $\hsa$ and $\hsd$ and they had formed separate classes in $K^1$, which is T-dual to $K^0(\rpt\times\rp^{2k+1}\times S^{6-2k})$.

% =========================================================================
\section{D3-brane Insertion on $\rp^1\times\rp^{2k-3}\times S^{6-2k}$} \label{inserzioni}

\subsection{The S-dual Configuration with Torsion $G_3$} \label{sduale}

While the twisted K-theories of all of the spaces above are well-defined, $\rpt\times\rp^{2k+1}\times S^{6-2k}$ configurations with torsion $H$-flux
\beq
H=ab^2+ba^2\,,
\eeq
do not correspond to S-covariant K-theory classes when $k>1$ because
\beq
Sq^3H=H\cup H =a^2b^4\neq 0\,.
\eeq
For example if the other fluxes $G_{2j+1}$ all vanish then, if we implement S-duality by simply interchanging the integral classes $G_3$ and $H$, we find that now the S-dual fields are
\beq
\tilde{H}=0\hsp \tilde{G}_3=ab^2+a^2b \,,
\eeq
and so there is a Freed-Witten anomaly
\beq
d_3 \tilde{G}_3=(Sq^3+\tilde{H}\cup )\tilde{G}_3=Sq^3\, \tilde{G}_3=a^2b^4\,. \label{fw}
\eeq
This is to be expected, compact flux configurations in the presence of branes do not correspond to K-theory classes because the fluxes are not closed under the AHSS differentials.  In this case the unit of D3 charge
\beq
Q_{\textup\footnotesize{\textup{D}}3}=d_3\tilde{G}_3=a^2b^4=1\in\H^6(\rpt\times\rp^{2k+1}\times S^{6-2k})=\Z_2^2\,,
\eeq
is dual to the nontrivial element of the $\Z_2$-valued homology cycle
\beq
\H_4(\rp^1\times\rp^{2k-3}\times S^{6-2k})\,.
\eeq
In short, this S-dual story is a typical application of the Freed-Witten anomaly.  In addition there may be additional D-branes required to cancel tadpoles of a purely gravitational origin, such as the D3-brane dual to $P$. 

However if $k=3$ then
\beq
\tilde{G}_3\cup \tilde{G}_3\cup \tilde{G}_3=a^3b^6\neq 0 \,,
\eeq
and so the D3-brane, which is Poincar\'e dual to $\tilde{G}_3\cup \tilde{G}_3$, itself supports $\tilde{G}_3$-flux equal to the torsion generator $ab^2$ of $H^3(\rp^1\times\rp^3)=\Z\oplus\Z_2$, where we have dropped the unimportant $S^0$ factor.   This is S-dual to a D3-brane that supports torsion $H$-flux, which leads to a Freed-Witten anomaly that forces the D3-brane to source a single D-string that wraps the dual $\rp^1\subset\rp^3$.  In this case the D3-brane worldvolume action term
\beq
S_{D3}\supset\int C_2\wedge *F \,,
\eeq
implies that a D3-brane supporting a unit of $\tilde{G}_3$-flux has a unit of electric charge on its worldvolume gauge theory.

The D3-brane is compact and so the total charge must be canceled, which requires the insertion of a fundamental string, whose endpoint on a D-brane is an electric charge.  The fundamental string, like its S-dual, wraps $\rp^1$ and extends away from the D3-brane.  However our total spacetime is compact and so this F-string must have two ends.  Yet there is no obvious candidate for another brane on which this string may end without creating another anomaly, since there are no nontrivial K-classes corresponding to the potential wrappings of the required 1 or 5-brane baryons.  Thus the configuration is apparently inconsistent once we take into account the S-dual Freed-Witten anomalies, although we cannot rule out the possibility that some more complicated anomaly allows this F-string to be absorbed.  For example if $P$ contains a term $a^2b^4$ then there will be a gravitational D3 wrapping the same cycle as the D3-brane above, which will cancel its D3-brane charge or equivalently supply a second endpoint for the string and so restore the consistency of the configuration. 

% =========================================================================
\subsection{Torsion $H$}

In the story of interest, however, it is not $G_3$ whose square is nontrivial but rather $H$.  This means 
that we need to replace Eq.~(\ref{fw}) with the S-dual Freed-Witten anomaly.  This is the Freed-Witten 
anomaly as derived for the worldvolume of an NS5-brane instead of a D5-brane.  The usual derivation of the 
anomaly relies upon the use of the worldsheet of a perturbative string which can not obviously be made 
BPS.  An application of S-duality to that argument may force us to go to a region of moduli space 
where the strong coupling effects render such perturbative strings unavailable, although in the large volume limit this calculation should nonetheless be reliable \cite{MW}.  And so rather than trying to S-dualize the argument of Freed and Witten, we will instead define the action of S-duality on torsion elements of integral cohomology by extending its action upon the elements of real cohomology $dB$ and $dC_k$.    That is, we will say that S-duality exchanges the integral classes $H$ and $G_3$.  We will then conjecture that IIB string theory is S-duality covariant and search for inconsistencies. 

When $k>1$ our configuration with torsion $H\neq 0$ and $G=0$ corresponds to a twisted K-theory class in $K^1_H$ but not to a class in the untwisted K-theory $K^1_{\tilde{H}=G_3=0}$ with torsion $\tilde{G_3}=H$ due to the anomaly (\ref{fw}).  In particular our configuration has a Freed-Witten anomaly which is S-dual to Eq.~(\ref{fw})
\beq
Q_{\textup\footnotesize\textup{D}3}=dG_5=d_3 H=H\cup H=a^2b^4 \,,\label{fwdual}
\eeq
leading again to a D3-brane supported on $\rp^1\times\rp^{2k-3}\times S^{6-2k}$.  Again we are not considering the possible gravitational correction $P$ to $Q_{\textup\footnotesize\textup{D}3}$.  D3-branes are invariant under S-duality, and according to our conjecture $\Z_2$-charged D3-branes are invariant too.  Thus we have found that the brane that cures this anomaly is just the S-dual of the brane that cured the S-dual anomaly above, which is not surprising as our formula for D3 charge is S-duality invariant.

As another check we can see what happens to the system when the $H$-flux is turned on or off.  An $H$-flux can be turned on or off by passing an NS5-brane that sources it.  For example in Ref.~\cite{MMS, EvsBranes} where twisted K-theory classified fluxes at moments in time, an NS5-brane that sweeps out a 6-cycle during some interval of time will change the $H$-flux on the dual cycle.  If the cycle contains $G_3$ or if its normal bundle is not $spin^c$ then the NS5-brane worldvolume will have an anomaly that is canceled by a D3-brane insertion, making a baryon \cite{baryons} configuration in which an NS5-brane sweeps out a cycle and a D3-brane ends on the NS5-brane and continues toward an infinity in the time direction.  Thus the initial and final conditions differ by a unit of $H$-flux and also possibly some D3-brane charge which is equal to the anomaly on the NS5-brane worldvolume and is approximated by Eq.~(\ref{fwdual}).   In fact such instantons interpolate between brane configurations that are homologically distinct but represent the same twisted K-homology classes in the S-dual twisted K-homology $K_{G_3}$.

In our case there is no time direction.  We have used all 10 directions to 
construct $\rpt\times\rp^{2k+1}\times S^{6-2k}$.  However we are not using K-theory to find the conserved 
charges on a timeslice.  Rather we are using K-theory to classify configurations on all of spacetime, as 
in Ref.~\cite{Witb}.  In that paper, rather than transforming between different cohomological 
representatives of the same K-class on different timeslices via instantons that act over an interval of 
time, the author uses Sen's construction \cite{Sen} to transform between cohomological representatives of 
the same K-class on the whole of spacetime via instantons that act over an interval in a deformation 
direction that is not one of the 10 dimensions of the spacetime.  The deformation direction in this case 
corresponds to the RG flow of the string field theory undergoing tachyon condensation.  Similarly the 
relation between different representatives of the same class of the S-dual twisted K-theory of the 
deformed conifold was seen to be an RG flow, the Klebanov-Strassler cascade, in Ref.~\cite{EvsMay}.  In 
short, the RG flow means that we can introduce an extra non-physical deformation dimension in which we may 
place Maldacena-Moore-Seiberg (MMS) instantons that transform between distinct cohomology representatives 
of the same 
twisted K-theory class.  The K-classes may then be identified with the universality classes of the theory whose flow we have used.  The deformation direction appeared similarly in Ref.~\cite{Hor}.

 Now we may toggle the H-flux by wrapping an NS5-brane around the oriented $\Z_2$-valued cycle of $\H_7(\rpt\times\rp^{2k+1}\times S^{6-2k})$ that came from the $Tor$ term in the K\"unneth formula.  This cycle is the Poincar\'e dual of $ab^2+ba^2$, which is just the dual of $ab^2$, $\rp^2\times\rp^{2k-1}\times S^{6-2k}$, glued to the dual of $a^2b$, $\rp^1\times\rp^{2k}\times S^{6-2k}$, along their common $\rp^1\times\rp^{2k-1}\times S^{6-2k}$, which is cut in each component to make the components orientable.   Note that the NS5-brane spans 7 dimensions, instead of the usual 6, in fact all branes extend in one extra dimension now that we have included the deformation direction.  However the NS5-brane does not extend in the deformation direction, which is the reason that as we pass the NS5-brane in the deformation direction the $H$-flux changes, toggling the $\Z_2\subset\H^3(\rpt\times\rp^{2k+1}\times S^{6-2k})$ on or off.  
 
When $k>1$ $H\cup H\neq 0$ implies that this NS5-brane is anomalous.  This is the reason that the flux that it sources suffers from a Freed-Witten anomaly.  The NS5-brane wraps a cycle whose normal bundle is not $spin^c$, and so it may only be rendered consistent if it also sources a $\Z_2$ charged D3-brane.  This 5-dimensional D3-brane ends with codimension 3 on the NS5-brane and continues to plus or minus infinity in the deformation direction.  Thus the D3-brane charge, which itself is $\Z_2$ valued, is toggled at the same time as the $H$-flux is toggled.  When $H$ vanishes the D3 charge needs to vanish (more generally it is equal to $P$) because D3 is a source for $dC_4$ which is gauge invariant in the absence of $H$-flux and so the integral of $ddC_4$, the total D3-brane charge, must vanish over the boundaryless $\rpt\times\rp^{2k-3}\times S^{6-2k}$ by Stokes' theorem.  If the D3-charge is zero when $H$ is zero and the D3-charge changes when $H$ changes then the D3-charge must be $1\in\Z_2\subset \H_4$ when $H$ is $1\in\Z_2\subset\H^3$.  In conclusion we have again used the S-dual Freed-Witten anomaly, this time on the worldvolume of the NS5-brane, and again we have found the same result, that our configuration with $H$ flux has a unit of D3-brane charge wrapped on $\rp^1\times\rp^{2k-3}\times S^{6-2k}$.

% =========================================================================
% Section 6
\section{T-Dualizing to D2} \label{inserzioniduale}

\subsection{The T-Duality}

We have found that $\Z_2$-valued D3-branes, one of which may be inconsistent, are required for anomaly cancellation in the cases $k=2$ and $k=3$.  These branes cannot be removed without also removing the $H$-flux.  However the T-duals, in particular the fluxless product of spheres, have no Freed-Witten anomalies and so the above restriction appears to have disappeared.  This means that during the course of T-dualizing the D3-branes need to disappear.  In this section we provide a proposal for how the brane charge might change.  Notice that if $P=a^2b^4$ then the branes are simply canceled by the gravitational branes before any dualities. 

In the case $k=3$ the D3-brane is a baryon which sources a D-string as it wraps 
nontrivial $H$-flux.  Its T-dual is therefore a kind of baryon as well, as the D-string dualizes to a D2-brane that ends on our dual D2.  The twisted K-theory classification does not include D-brane-D-brane baryons \cite{EvsBranes} and so this brane does not dualize to a class in twisted K-theory.  We do not need to use the claim that the compactification is inconsistent in this argument.  However the fact that the baryon has a brane of the same dimension ending it suggests the seriousness of the topological obstruction confronting it.  In fact it wraps a hemisphere that has a boundary, and so the D2-brane ending on it is the brane continuing out from that boundary.

The case $k=2$ is more interesting.  Although $\rp^5$ is not $spin$, the T-duality to $S^3\times S^5\times S^2$ without fluxes ensures that the spectrum of the full string theory is supersymmetric, as in Ref.~\cite{DLP}. As noted in the introduction, since the dual spheres are small the T-duality does not prove that string theory can be defined on this non-$spin$ space.\footnote{The same applies the example of Ref.~\cite{DLP} and so we are optimistic that, using the $H$ flux, it can be defined.}  If it cannot then we need to consider the $\rp^5$ to be an orientifold.  This would change the classification of fluxes.  For example, if we replace the $S^2$ by a $T^2$ then the desired orientifolding may be achieved with an O3-plane that wraps the $\rpt$ and a circle in the torus.  In this case the three and seven-dimensional field strengths will be classified by $\Z_2$-twisted cohomology \cite{baryons,HK} and in particular will be $\Z_2$-valued.  The five class, under which our D3 is charged, will still be classified by integral cohomology.  However it is no longer clear that our $H$-flux is coclosed.  We will not consider this case further. 

While we cannot prove that the twisted $k=2$ string theory exists, we can still study the charges and duality transformations that the D3-brane will have if it indeed does exist.  The D3-brane wraps $\rp^1\times\rp^1\times S^2\subset \rp^3\times\rp^5\times S^2$.  Recall that we constructed $H=ab^2+a^2b$, by taking the Bockstein of a two-class $ab$.  Thus we identify the 2-class $ab$ with the B-field and use the homology-cohomology pairing to write
\beq
\int_{\rp^1\times\rp^1}B\sim 1\,. \label{tauro}
\eeq
The integral of the $B$-field over a 2-cycle of a D$p$-brane measures the 
worldvolume D$(p-2)$-charge, and so already we see that this D3-brane carries D1-brane charge.  This is a consequence of the extension problem, in which the corresponding
\beq
\Z_2=\H^2(\rpt)\otimes \H^4(\rp^5)\subset \H^6(\rpt\times\rp^5)\,,
\eeq
was absorbed by 
\beq
\Z=\H^3(\rpt)\otimes \H^5(\rp^5)=\H^8(\rpt\times\rp^5)\,,
\eeq
via the extension
\beq
\H^8(\rpt\times\rp^5)=\Z\longrightarrow\Z\longrightarrow \Z_2\subset \H^6(\rpt\times\rp^5)\,.
\eeq
In terms of cohomology classes the extension problem set
\beq
2a^2b^4=a^3b^5 \,,
\eeq
which, after Poincar\'e dualizing, means that our D3-brane carries half a unit of charge of D1-brane wrapped around the $S^2$.
Thus we see that while our D3-brane appears to be torsion, even multiples of the D3 do not decay into nothing, but rather they leave D-strings that wrap the 2-sphere.  This sets the normalization of the $B$-flux in Eq.~(\ref{tauro}) to one half.

If $P=a^2b^4$ then there will be a second D3 wrapping the same cycle, and so the two D3's will annihilate.  The nature of the gravitational anomaly that the D3 cancels will determine its worldvolume D1 charge, which will be half-integral.  Depending on the D1 charges carried by the D3-branes, which are in principle determined entirely by the anomalies of the configuration, there will be some integral number of D1-branes remaining after the annihilation.

Although the T-dual has no torsion homology, we may still see the extension problem in action in the dual picture.  We have seen (\ref{tauro}) that the D3-brane wraps a torus $\rp^1_a\times\rp^1_b$ that supports a half unit of NS $B$-flux, while its worldvolume $U(1)$ gauge field strength $F$ is quantized.  $B+F$ is gauge-invariant and we may formally construct a pseudo-bundle that has Chern class $B+F$.  The connection of this bundle $A$ may be integrated over $\rp^1_a$ at various values of $\phi_b\in\rp^1_b$ to yield a Wilson loop
\beq
f(\phi_b)=\int_{\rp^1_a\times\phi_b}A\,.
\eeq
As $B+F$ is half-integral, the Wilson loop is not single valued, but rather it shifts by $\pi$ each time one encircles $\rp^1_b$
\beq
f(\phi_b+2\pi)=f(\phi_b)+\pi\,.
\eeq

The D3-brane wraps the $\rp^1_a$ which is T-dualized, and so it is dual to a D2-brane that is localized at a point $\hat{\phi}_a$ on the dual circle $\hsa$.  This point is determined by the Wilson loop
\beq
\hat{\phi}_{a}(\phi_b)=f(\phi_b)\,,
\eeq
and so each time one encircles $\rp^1_b$, the brane goes half way around $\hsa$.  This is just the construction of $\hsd$, which is half of $\hsa$ plus $\rp^1_b$, and so the dual D2-brane wraps $S^2$ times the $\hsd$, which is the generator
\beq
\hsd=1\in \H_1(S^2\times S^5\times S^1)\,.
\eeq
Meanwhile the dual of the D1-brane that wrapped the $S^2$ is a D2-brane that wraps $S^2\times\hsa$.  This is because it did not wrap the original circle $\rp^1_a$ so it must wrap the dual circle $\hsa$.  However $\hsa$ corresponds to the element
\beq
\hsa=2\in H_1(S^2\times S^5\times S^1) \,,
\eeq
and so, as indicated by the extension problem, the D1-brane (which generated $\Z$) dualizes to twice the D3-brane (which generated $\Z_2$).  If $P=a^2b^4$ then the D2 will wrap $\hsa$ an integer number of times, or equivalently it will wrap $\hsd$ an even number of times.

% =========================================================================
\subsection{The D2 Disappears When a Global $B$-Field is Included}

It seems a bit strange that the $D2$-brane disappears when a global 
$B$-field is included.
After T-dualizing in the case $k=2$ we have found that the K-class corresponding to D1-branes on the $S^2$ is odd.  No such restriction exists on the product of spheres $S^2\times S^5\times S^1\times S^2$ with no fluxes that appears after T-dualizing $\hsd$, and so as a consistency check we will investigate how this condition might disappear.  In the process we find a new variant of the Freed-Witten anomaly.

After the first T-duality the $H$-flux is
\beq
H=2\a\cup \theta_a \,,
\eeq
where $\theta_a=d\phi_a$ generates $H^1(\hsa)=\Z$.  To perform the second
T-duality, along $\hsd$, we claim that we need to\footnote{Otherwise we might expect to find something in the metric and the $B$-field of the dual product of circles $S^3\times S^5\times S^2$ that produces D1 charge on the $S^2$, such as a twisted version of an S-dual of the 1-loop effect in Ref.~\cite{VW}.  If no such effect exists then this would suggest that $P=a^2b^4$.}  decompose $H$ into a
part along and transverse to $\hsd$.  That is, we wish to write $H$ in
terms of $\theta_d$ instead of $\theta_a$.  This is no problem
topologically, as $2\theta_a$ is cohomologous to $\theta_d$, and for
the twisted K-theory automorphism only the cohomology class of $H$ is
important \cite{RR, BEMa}.  In particular we may rewrite $H$ as
\beq
H\p=\a\cup \theta_d\,.
\eeq
However geometrically ${H}$ and ${H\p}$ differ by an exact form $B$
\beq
\Delta H=H-H\p=\a\cup (2\theta_a-\theta_d)=d((2\phi_a-\phi_d)\a)=dB\,.
\eeq
$\hsa$ is twice as long as $\hsd$, and so the function $2\phi_a-\phi_d$ is a stepfunction which is zero the first time around $\hsd$ and then $2\pi$ the second.  While $\Delta H$ is exact and so topologically trivial, we will see that it enters an anomaly multiplied by a gauge-dependent term and the product is topologically nontrivial.

To see this we perform a 9-11 flip (this is the same as T-dualizing
the S-dual description of Sec.~\ref{sduale} in which ${\tilde{G}}_3\neq 0$ and ${\tilde{H}}=0$) so that $\hsa$ is the M-theory circle.  Now our IIA spacetime is $\cp^1\times\rp^5\times S^1_m\times S^2$ where $S^1_m$ is the former M-theory circle.  The new $H$ is roughly
\beq
H=2\a\cup \theta_m\,.
\eeq
However the M-theory lift of $\Delta H$, which is $\Delta G_4$, is not independent of the M-theory coordinate $\phi_a$.  On the contrary since it is the derivative of the step function it is a Dirac delta function at $\phi_a=0$ and $\phi_a=\pi$, with opposite signs at the two values.  

Now there is also $G_2$-flux
\beq
G_2=b^2\,.
\eeq
We recall, from the extension problem for the K-theory of $\rp^{2k+1}$, that $G_2$-flux carries half a unit of $G_4$-flux.  That is to say
\beq
2G_4=sq^2 \,G_2= G_2\cup G_2 = b^4\,.
\eeq
Now we use the Freed-Witten anomaly
\beq
{\text{PD}}(\textup{D}2)=ddC_5=\Delta H\cup G_4=\a\cup \theta_m\cup b^4\,,
\eeq
to conclude that the shift in $H$ by an exact form yields a half D2-brane wrapping $\rp^1\times S^2$ at $\phi_a=0$ and an anti half D2 at $\phi_a=\pi$, the two points where $H$ is nonzero.  As $\H_3(\rp^1\times S^2)=\Z_2$ there is no topological difference between a D2 and an anti-D2 wrapped on this cycle but the orientation may mean that they carry opposite F-string charges, which are $\Z$-valued.  The D2-brane worldvolume coupling
\beq
S_{\textup\footnotesize\textup{D}2}\supset\int C_1\cup B\,,
\eeq
combined with
\beq
1=G_2=d C_1\,,
\eeq
implies that half D2-branes each carry half-integral charge of F-string wrapped
around $S^2$.  As always, this F-string is an M2-brane and wraps the
M-theory circle, $\hsa$.  However the relative signs of these two half-strings depend on a lift of their $\Z_2$ classes that we have not determined and perhaps it cannot be determined without finding additional consistency conditions satisfied by the strings, although it seems plausible that the lifts of a brane and antibrane cancel, suggesting $P=a^2b^4$.  Thus the total F-string charge changes by an amount that depends on their unknown lift.    Now doing a 9-11 flip back we find that each F-string becomes a D2-brane that wraps $S^2\times\hsd$, which as required is the T-dual of the D3-brane insertion in IIB.  If we can calculate the above lift then we will be able to compute change in D-brane charge, which will in turn allow us to compute $P$.

Thus we have learned that the addition of an exact form to $H$ can potentially, due to an S-dual composite Freed-Witten anomaly, toggle the D2-brane charge that is produced by our T-duality.  As this addition of the exact form is apparently necessary between the two T-dualities if we wish to use the T-duality prescription of \cite{BEMa,BEMb}, we find that the charge of our D3 insertion may be canceled before we arrive at $S^3\times S^5\times S^2$ with no flux.  We feel that this result, despite our inability to compute the lifts, teaches us about the limitations of the K-theory program.  The K-theory classification relies heavily on forgetting the Wilson loops.  But here we see that the Wilson loops may be able to affect the D-brane charges.

It may seem as though we have only replaced one problem with an equivalent problem.  After T-dualizing from $\rp^3\times\rp^7\times S^2$ we found that the original D-brane insertion implied that the brane charge is, up to gravitational corrections, an odd element of the $\Z=K_1(\cp^1\times S^5\times S^1_d\times S^2)$.  This did not seem to agree with the physics of the final spacetime $S^3\times S^5\times S^2$, but we found that before doing the second T-duality we need to include a globally defined B-field that can change our $K_1$ class by one unit in $\Z$.  Now the K-homology class is even instead of odd, and so we may ask again how this restriction arises from the viewpoint of the product of spheres.  The answer, as was explained in the introduction, is that because our spacetime is compact the only allowed K-homology class for D-brane wrappings is zero\footnote{More precisely we argued that it is the class for which there are only Hanany-Witten type brane charges.  Thus the allowed class is odd before we change the B-field, 0 after, and 0 on the final $S^3\times S^5\times S^2$.}, which is even and so there is no contradiction.  To get branes filling out the entire twisted K-homology we may, for example, replace the 2-sphere with a noncompact space.  In this case the K-class of the product of spheres may be either odd or even.  Correspondingly, the original product of projective spaces may now have a $dF$ type D3-brane that wraps $\rp^1\times\rp^1\subset \rpt\times\rp^5$ and cancels the D3 charge but preserves the consistency of the configuration.  Thus in the noncompact case there may be any net D3 charge in the initial configuration and so the K-class may be odd or even in the T-dual configuration, while in the compact case the original D3-charge is 1 corresponding to a final K-class which is 0.  In both cases we have then seen that the allowed charges before and after the T-duality can agree, as they must.

% =========================================================================
\subsection{A Correction to the $W_7$ anomaly}

The anomaly used in this argument may be summarized as
\beq
{\text{PD}}(\textup{F-String})=ddB_6=H\cup G_2\cup G_2\cup C_1\,.
\eeq
At an intermediate step, to construct the D2-brane, we used a simpler anomaly
\beq
{\text{PD}}(\textup{D2})=ddC_5=H\cup G_2\cup G_2\,.
\eeq
This appears to be an extra term in the Diaconescu-Moore-Witten (DMW) anomaly \cite{DMW}
\beq
W_7=0. \label{anom}
\eeq
Eq.~(\ref{anom}) is a special case of the Freed-Witten anomaly
\beq
{\text{PD}}(\textup{D2})=ddC_5=(Sq^3+H\cup)G_4 \,,
\eeq
because when $H$ vanishes and there are no D2-branes the flux quantization condition on spin manifolds
\beq
G_4=w_4\ \textup{mod}\ 2 \,,\label{quant}
\eeq
yields the DMW anomaly
\beq
W_7=Sq^3(w_4)=Sq^3 G_4=0\,.
\eeq
The flux quantization condition (\ref{quant}) has been demonstrated on $spin$ manifolds, and the current spacetime is not $spin$.  For non-$spin$ manifolds counterexamples are known 
\cite{DLP, BEMa}.  In this case in fact it holds because
\beq
w_4(\rp^5)=\left( \begin{matrix} 6 \\ 4 \end{matrix} \right)\mod 2=1\,,
\eeq
and $G_4$ is nontrivial as $G_2$ produces a half unit of $G_4$-flux via $G_2\cup G_2$.  Thus the applicable quantization condition in this non-$spin$ case may include a contribution
\beq
G_2\cup G_2\ \textup{mod}\ 2  \,,
\eeq
which (since $Sq^3G_2=0$) leads to a contribution to the D2 charge of
$H\cup G_2\cup G_2$.  Combining these contributions suggests that the
mod $2$ part of the D2-charge may need a correction.  For example when the
M-theory manifold is $spin$ then $G_2=w_2$ mod 2 and so one may expect
an $H\cup w_2\cup w_2$ contribution.

% =========================================================================

\section{Open Questions}

Perhaps the most surprising feature of the above examples is
the appearance of extra factors of one half in the normalization of some
D-brane charges.  Even in the case of $\rps\times S^3$ this one half
meant that a D-brane wrapping $\rpt$, which has a $spin$ normal bundle
and no $B$-flux, contains half a unit of $\rp^1$-wrapping D-brane charge.  This $\rp^1$ brane corresponds to
a bundle with all vanishing Chern classes, but the corresponding term
in the spectral sequence was nontrivial. Perhaps one may reconstruct
the nontrivial spectral sequence element by combining the Chern class
and the $\sqrt{\hat{A}}$ terms of the D-brane effective action in such
a way as to preserve this torsion.  After all the worldvolume action
of the brane does presumably contain the information about lower
dimensional brane charges.  The other possibility is that the factor of
one half is always present, in which case one needs to learn how to
make this division by two canonical.

Another puzzling fact is the inclusion of the half D2-brane charge in
the $\Z$-valued K-theory.  The spacetime is compact and so the
integral of $dG_6$ must be zero and so the D2 charge is precisely
determined by the Freed-Witten type terms.  The D2 charge is T-dual to
the $\Z_2$ valued D3 charge and so is odd, but different odd values
are related by $dG_6$ charged D2 branes, and so only one odd value is
consistent.  One may then ask what determines this value, that is,
what does the nontrivial element of $\Z_2$ lift to once the $\Z_2$ is
included in $\Z$ by the extension problem.  This is the same as
choosing the integral part of $B$.  One may ask whether this choice
is merely a choice of gauge, or an observable.  But if it is
observable then conceivably it corresponds to a choice on the original
product of projective spaces, where there is not obviously any
$\Z$-valued ambiguity.  Even multiples of D2's are dual to D1's in the original picture, and so it is reasonable to conjecture that the number of pairs of D2's is determined by whatever consistency condition determines the D1 charge.  Again it would be useful to find a formula for D1 charge like the formula that we propose for D3 charge.

In \cite{KS1} it was shown that the DMW anomaly,
given by $W_7$, is absent if the partition function is defined in 
(complex oriented) elliptic cohomology. The study was done for the 
partition function of the fields in the absence of branes, in the
same spirit as DMW. In the current paper we chose to include D-branes
in the analysis and that has led us to propose a modification of
the $W_7$ condition due to the presence of D2-branes. It would be 
interesting to see how the corresponding discussion in \cite{KS1} 
would be extended. 

One main question is to what extent S-duality is compatible with 
(twisted) K-theory. In \cite{KS2} this was studied starting from 
the conjecture in \cite{DMW}, proven in \cite{DFM}, that in the absence of D3-brane charge
\begin{equation}
H\cup H + G_3\cup H + G_3 \cup G_3 +P =0\,. 
\label{dmwe}
\end{equation}
In this context, it was shown in \cite{KS2} that the $H\cup H$ 
term causes an affine twisting, but subsequently that it is 
inconsistent in the framework of any $K(\Z,2)$-twisting. Further,
the $P$ term was also shown to cause affine twisting implying that 
in order to have S-duality in type IIB in ten dimensions, 
twistings by $H$ must be accompanied by some higher-dimensional 
non-trivial twisting. Such higher twistings, as pointed out in \cite{KS2}, 
as 
well as the P term,
in fact correspond to constructing a Postnikov tower of a classifying 
space. In \cite{KS2}, it was proposed that this space should correspond to 
a generalized cohomology theory, which was conjectured
to be a form
of elliptic cohomology. It is, however, alternately possible to 
construct this Postnikov tower directly, identifying all the homotopy
groups (such as P and higher twistings), and the Postnikov invariants
between them, which correspond to equations such as (\ref{dmwe}).
More on this will appear in the future.
However, it is also evident from that work that in looking at the
problem from a higher-dimensional perspective, e.g. in twelve dimensions, 
one seems to inevitably need elliptic cohomology. (Some aspects of
elliptic cohomology in twelve dimensions has
been discussed recently in \cite{KS3}.)

There are a number of directions for future research.  For example
this approach may be used to try to construct the Atiyah-Hirzebruch
differentials in general.  While S-dual anomalies require that we
restrict attention to 10-dimensions, we found a term in $d_5$ in the
untwisted case where the S-dual effects were not important.  In fact
it seems as though there is a consistent truncation of the formalism
in which the S-dual gauge transformations and S-dual Freed-Witten anomalies
are ignored, so that one may consider more dimensions and thus use the
worldvolume D-brane charges to construct the higher differentials.
This truncation appears automatically in the context of
conformal field theories and so this approach corresponds to the fact that we may consider conformal
field theories with targets of dimension greater than 10 describing
open strings whose possible boundary conditions will be classified by the
desired K-classes.

If we go beyond this approximation then string theory provides
modifications of K-theory which still have not been identified.  For
example the inclusion of S-duality seems to lead to an infinite family
of K-theories that are related by a set of $SL(2,\Z)$
transformations.  The full supergravity also has the relations
\beq
G_p=*G_{10-p} \,,
\eeq
where the Hodge dual $*$ is generally irrational and so generically at
most half of the RR field strengths are integral at a time.  An analogy
with other systems, such as the chiral scalar in 2 dimensions, suggests that this
means we need to quantize K-theory such that only half of the Chern
characters are defined at a time.  Proposals for these two variations
of K-theory will appear elsewhere.

% =========================================================================
               
\section* {Acknowledgements} 

We are very grateful to G. Moore and E. Witten for extensive help with this 
project and also to C. Douglas and H. Miller for explaining the role of 
secondary cohomology operations. P.B., V.M. and H.S. are supported, in part, 
by the Australian Research Council, and
J.E. is partially supported by IISN - Belgium (convention 4.4505.86), by
the ``Interuniversity Attraction Poles Program -- Belgian Science
Policy'' and by the European Commission FP6 programme
MRTN-CT-2004-005104, in which he is associated to V. U. Brussel.  
B.J. is partially supported by the European Commission 
RTN program MRTN-CT-2004-005104. J.E. and B.J. would like to thank the 
University of Adelaide for hospitality while this work was in progress. B.J. would
also like to thank the INFN Torino for hospitality and JE would 
like to thank Harvard University. H.S. would like to thank the Shanghai 
Institute for Advanced Studies and the
University of Science and Technology of China for 
hospitality during the intermediate stages of this project.

% JE would like to thank MW for discussions?

% \newpage

% =========================================================================
 \section* {Appendix}
 
The real projective space $\rp^n$ is the quotient space of the sphere
$S^n$ under the action of $\Z_2$ generated by antipodal maps 
$x\mapsto -x$. In particular, $\rp^1$ is just the circle $S^1$.
Since each hemisphere in $S^n$ is disjoint from its antipodal image,
the $\Z_2$ action is a covering space action. Since $S^n$
is simply connected for $n\geq 2$, the covering  
$S^n \rightarrow \rp^n$  gives the fundamental group
$\pi_1(\rp^n) = \Z_2$ for $n \geq 2$. The generator of
this group is any loop obtained by projecting a path in
$S^n$ connecting two antipodal points.
As a consequence, one has the first homology group
$\H_1(\rp^n) = \Z_2$.

Let $L=S^n \times \R/\Z_2$ be the real classifying line bundle over
$\rp^n$ and let $x=w_1(L)$ generate $\H^1(\rp^n; \Z_2)=\Z_2$. The tangent 
bundle of $\rp^n$ is 
\begin{equation*}
T(\rp^n) \oplus 1=(n+1) L \,.
\end{equation*}

The Stiefel-Whitney classes are given as follows. Let 
\begin{equation*}
\left( \begin{matrix} n \\ i \end{matrix} \right)_2 = \frac{n!}{i! (n-i)!}\quad \mod2\,,
\end{equation*}
be the binomial coefficient reduced modulo $2$
(since we are dealing with the Steenrod algebra).
Then the formula for the Stiefel-Whitney classes for $\rp^n$ is
\bea
w_i(\rp^n)=\left( \begin{matrix} n+1\\
i \end{matrix} \right)_2 x^i\,.
\nonumber
\eea

Of interest are the first and the second Stiefel-Whitney classes,
which characterize whether the manifold is orientable and {\it spin}
respectively. So for $i=1,2$, we have
\bea
w_1(\rp^n)&=&(n+1) x\,,
\nonumber\\
w_2(\rp^n)&=&\frac{1}{2}n(n+1) x^2\,,
\nonumber
\eea
keeping in mind that the coefficients are taken modulo two. 
In order for the manifold to be orientable, $w_1$ has to vanish,
which implies that $n$ must be odd, as it is always in this note.
Next, the {\it spin} condition is that both $w_1$ and $w_2$ vanish. This
implies that $n=4k+3$. In particular, $\rp^3$ and $\rp^7$, which we
use, are {\it spin} manifolds. 
In this case one can calculate the first cohomology with $\Z_2$
coefficients, $\H^1(\rp^{4k+3}, \Z_2)$, to be $\Z_2$, which implies
that there are two inequivalent spin structures on $\rp^3$ and 
$\rp^7$. Of course we also know that $\rp^1=S^1$ which is {\it spin} and also
has two spin structures (Ramond and Neveu-Schwarz).
What about $\rp^{4k+1}$? Again, inspecting the formulae, one notices 
that in this case $w_1$ vanishes but $w_2$ is the reduction of
an integral class. This is the ${\it spin}^c$ condition. In particular,
in this paper we used the fact that $\rp^5$ is ${\it spin}^c$.

In the text we are interested in products of projective spaces.
 Note that the product of two orientable manifolds is also orientable,
the product of two {\it spin} manifolds is also {\it spin}, and the
product of two ${\it spin}^c$ manifolds is also ${\it spin}^c$.

One can also see that, besides the question of spin, there are other 
differences between $\rp^5$ and $\rp^n$ for $n=1,3,7$. One can see
such a difference in the context of (complex) K-theory. As above, if 
$L_{\C}$ is the corresponding complex line bundle, then one has
$\left( T(\rp^n) \oplus 1\right)\otimes \C = 2k \cdot L_{\C}$, where $n=2k-1$. 
The reduced K-theory ${\widetilde K}(\rp^n)$ is a cyclic group of
order $2^{k-1}$ which is generated by $x=[L_{\C}]-1$. Thus the bundle
$T(\rp^n) \oplus 1$ is trivial if $2^{k-1}$ divides $2k$. This 
implies that $k=1,2,4$ or equivalently that $n=1,3,7$. Therefore
we have the important result that $\rp^1$, $\rp^3$ and $\rp^7$ are
parallelizable, the same way that their double covers $S^1$, $S^3$
and $S^7$ are.  This result can also be deduced from the quaternion 
and octonion multiplication. An important consequence of this is
that all their characteristic classes are zero.

% \newpage

% =========================================================================

% =========================================================================

\end{document}